\documentclass[prd,,aps,showpacs,nofootinbib,superscriptaddress,amsmath,amssymb]{revtex4}
\usepackage{amssymb,epsfig}
\usepackage{mathrsfs}
\usepackage[usenames,dvipsnames]{color}
\usepackage[bookmarks]{hyperref}
\usepackage[usenames,dvipsnames]{color}
\usepackage{multirow, array} %para las tablas
\usepackage{float} % para usar [H]

\newcommand{\Natural}{\mathbb{N}}
\newcommand{\Integer}{\mathbb{Z}}
\newcommand{\Real}{\mathbb{R}}
\newcommand{\Complex}{\mathbb{C}}

\newcommand{\sign}{\mbox{sign}}

\newcommand{\proof}{\noindent {\bf Proof. }}

\newcommand{\qed}{\hfill \fbox{} \vspace{.3cm}}

\newtheorem{lemma}{Lemma}

\newtheorem{theorem}{Theorem}

\begin{document}
%%%%%%%%%%%%%%%%%%%%%%%%%%%%%%%

\title{Phase space mixing in an external gravitational central potential}

\author{Paola Rioseco}
\affiliation{Instituto de F\'isica y Matem\'aticas,
Universidad Michoacana de San Nicol\'as de Hidalgo,
Edificio C-3, Ciudad Universitaria, 58040 Morelia, Michoac\'an, M\'exico}
\affiliation{Departamento de F\'isica y Astronom\'ia, Facultad de Ciencias, Universidad de La Serena, Avenida Cisternas 1200, La Serena, Chile}

\author{Olivier Sarbach}
\affiliation{Instituto de F\'isica y Matem\'aticas,
Universidad Michoacana de San Nicol\'as de Hidalgo,
Edificio C-3, Ciudad Universitaria, 58040 Morelia, Michoac\'an, M\'exico}

\begin{abstract}
This article is devoted to the study of the dynamical behavior of a collisionless kinetic gas in $d=1,2,3$ space dimensions which is trapped in a rotationally symmetric potential well. Although at the microscopic level the trajectories of individual gas particles are quasi-periodic and characterized by their $d$ fundamental frequencies, at the macroscopic level the gas relaxes in time to a stationary state, provided the potential satisfies a certain non-degeneracy condition.

In this article, we provide a mathematically precise formulation for this relaxation process which is due to phase space mixing. In particular, we prove that a physically relevant class of macroscopic observables computed from the one-particle distribution function, such as particle and energy densities, pressure and stress tensors, converge in time to the corresponding observables associated with an averaged distribution function. The latter can be determined from the initial datum and depends only on integrals of motion. Thus, the final state of the gas is described by an effective distribution function depending only on integrals of motion, which considerably reduces the degrees of freedom of the gas configuration. We discuss some applications to gravitational physics, including the propagation of a collisionless gas in typical potentials arising in stellar dynamics and the modeling of dark matter halos, and we also generalize our results to a relativistic gas whose individual particles follow bound timelike trajectories in the exterior region of a static, spherically symmetric black hole spacetime.
\end{abstract}

\date{\today}

% 04.20.-q: Classical GR
% 04.20.Ex: Initial value problem, existence and uniqueness of solutions
% 04.25.-g: Approximation methods; equations of motion
% 04.25.D-: Numerical relativity
% 04.25.Nx: Post-Newtonian approximation; perturbation theory; related approximations
% 04.40.-b: Self-gravitating systems, continuous media and classical fields in curved spacetime
% 04.70.-s: Physics of black holes
% 05.20.Dd: Kinetic theory
% 97.60.Lf: Astronomy: Late stage of star evolution: black holes

\maketitle

%%%%%%%%%%%%%%%%%%%%%%%%%%%%%%%%%%%%%%%%%%%%%
\section{Introduction}
%%%%%%%%%%%%%%%%%%%%%%%%%%%%%%%%%%%%%%%%%%%%%

One of the most fascinating subjects in physics is the emergence of irreversible macroscopic phenomena from reversible microscopic laws. In classical statistical mechanics, the passage from the microscopic to the macroscopic description is achieved by computing phase space averages. Denoting by $\Gamma$ the $2n$-dimensional phase space (with $n$ the number of degrees of freedom) and by $\rho_t:\Gamma\to \Real$ the ensemble density at time $t$ (normalized such that $\int_\Gamma \rho_t(q,p) d^n q d^n p = 1$), the average of an observable $O:\Gamma\to \Real$ at time $t$ is 
\begin{equation}
\langle O \rangle_t := \int\limits_{\Gamma} O(q,p) \rho_t(q,p) d^n q d^n p.
\label{Eq:AAverage}
\end{equation}
If the microscopic laws are described by an autonomous Hamiltonian system, inducing a phase flow $\varphi^t: \Gamma\to \Gamma$, then according to Liouville's theorem the evolution of the ensemble density is given by $\rho_t(q,p) = \rho_0(\varphi^{-t}(q,p))$ with $\rho_0$ denoting the density at time $t = 0$. A key question is whether the macroscopic averages~(\ref{Eq:AAverage}) reach an equilibrium state. This means there should exist an ensemble density $\rho_\infty: \Gamma\to \Real$ (depending on $\rho_0$ but not on $O$) such that
\begin{equation}
\lim\limits_{t\to\infty} \langle O \rangle_t 
 = \lim\limits_{t\to\infty} \int\limits_\Gamma O(q,p) \rho_0(\varphi^{-t}(q,p)) d^n q d^n p
 = \int\limits_\Gamma O(q,p) \rho_\infty(q,p) d^n q d^n p
\label{Eq:Mixing}
\end{equation}
for all observables $O$. This is the definition of ``mixing'' we shall adopt in this article. 

Note that in general, the pointwise limit $\rho_t(q,p)\to \rho_\infty(q,p)$ (for fixed $(q,p)\in \Gamma$) does not exist; hence one cannot pass the limit below the integral and determine $\rho_\infty$ in this way. Note also that $\rho_\infty$ depends on the energy $E$ since the motion is restricted to the surfaces of constant energy $H(q,p) = E$ with $H$ the Hamiltonian of the system. The strongest concept of mixing (sometimes referred to as ``chaotic mixing") occurs when Eq.~(\ref{Eq:Mixing}) holds with $\rho_\infty$ being constant on each energy surface and determined by averaging $\rho_0$ over each energy surface. This happens when the flow is sensitive to the initial conditions, such that two initial points in phase space which lie very close to each other become widely separated as time progresses (hence the name ``chaotic"). For precise mathematical definitions of this concept of mixing and its relation with ergodicity, we refer the reader to Refs.~\cite{jLoP73,CornfeldFominSinai-Book}.

The concept of mixing that is relevant for the present article is ``phase space mixing", which occurs for Hamiltonian systems which are integrable. For such systems, the motion is restricted to $n$-dimensional tori defined by $n$ integrals of motion (the action variables). Phase space mixing consists in showing that Eq.~(\ref{Eq:Mixing}) holds with $\rho_\infty$ being constant on each torus, where the constant is determined by averaging $\rho_0$ over the torus. It follows that the equilibrium ensemble density $\rho_\infty$ is a function depending only on the action variables (whereas the initial ensemble density $\rho_0$ is an arbitrary function on phase space $\Gamma$), such that the number of degrees of freedom is reduced from $2n$ to $n$. This phenomenon is expected to persist for sufficiently small perturbations of integrable Hamiltonian systems, for which most trajectories are still confined to invariant tori. In some situations (including examples studied in this article) there exist more than $n$ isolating integrals of motion,\footnote{See Ref.~\cite{dL62a} for a definition of isolating integrals and their role for Jeans theorem.} and in this case phase mixing takes place on lower-dimensional invariant subspaces and the equilibrium density depends on more than $n$ variables.

Phase space mixing plays an important role in many areas of physics, and hence it has been studied extensively in the literature. In the context of stellar dynamics in a galaxy, for instance, mixing has been invoked~\cite{dL62,dL67} to provide an explanation for the approach to equilibrium, even though interactions between the individual stars are negligible compared to the mean gravitational field. Further important work on the role of mixing in stellar dynamics can be found in Refs.~\cite{sTmHdL86,sT99,gCrSmFbGpKpA14}, see also section~5 in Ref.~\cite{dM99} for a review on collisionless relaxation in elliptic galaxies. Of particular interest to this article is the recent work in Ref.~\cite{pDeJmAeMdN17} which performs a numerical study of the Vlasov equation in a fixed background potential $\Phi$ motivated by well-known models for dark matter halos. In that work, it is observed that the one-particle distribution function relaxes in time to a stationary, virialized state.\footnote{See Ref.~\cite{BinneyTremaine-Book} for a formulation of the virial theorem in the context of kinetic theory.} In general relativity, mixing has been observed in numerical studies of the critical collapse of the spherically symmetric Vlasov-Einstein equations~\cite{aAmC14} and discussed in our previous work~\cite{pRoS18} regarding thin disks in the equatorial plane of a Kerr black hole.

In plasma physics, mixing turns out to be a key ingredient for the Landau damping effect, in which a charged collisionless gas relaxes in time to a homogenous configuration. Rigorous results regarding this effect have been given by Mouhot and Villani~\cite{cMcV11}, who proved phase space mixing for the full Vlasov-Poisson equations on a finite box with periodic boundary conditions without linearization. For an extension of these results to the special-relativistic generalization of the Vlasov-Poisson equations, see Ref.~\cite{bY16}. Phase space mixing also has interesting applications in quantum mechanics and field theory, see for instance~\cite{rMeT17,tDaKeKyS02,tDaKnRyS02}.

In this article, we analyze phase space mixing in the context of a collisionless kinetic gas which is trapped in a potential well defined by a rotationally-symmetric external potential $\Phi$ in $d$ space dimensions. In this case, the role of the ensemble density $\rho_t$ is played by the one-particle distribution function, $\Gamma$ is the $2d$-dimensional one-particle phase space, and the relevant integrals of motion are the energy $E$ and the angular momentum (if $d > 1$). The central result of this article is to provide a simple condition (referred to as the ``non-degeneracy condition" in the following) which is sufficient for phase space mixing~(\ref{Eq:Mixing}) to take place in such systems. For $d > 1$ this condition requires the Hessian of the area function $A(E,L)$ to be invertible for almost all values of the energy $E$ and total angular momentum $L$ of the orbits, where $A(E,L)$ is defined as the area enclosed by the orbit in the two-dimensional phase space $(r,p_r)$ describing the radial motion. (For $d=1$ this condition reduces to the well-known requirement for the period $T(E)$ associated with the orbit of energy $E$ to be nowhere constant.) As an application, we provide several examples of potentials $\Phi$ from stellar dynamics and the modeling of dark mater halos and show that they satisfy the non-degeneracy condition for mixing to take place in $d=1,2,3$ dimensions. In particular, our results show that the virialization process observed in~\cite{pDeJmAeMdN17} is due to phase space mixing. In addition, we generalize our results to the case of a general relativistic kinetic gas propagating on a static, spherically symmetric black hole spacetime, and show that mixing takes place provided (a relativistic generalization of) the area function $A(E,L)$ satisfies a non-degeneracy condition which is analogous to the Newtonian case.

The remainder of this article is organized as follows. In section~\ref{Sec:Main} we describe the mathematical models for a collisionless gas in an exterior Newtonian potential in arbitrary dimensions, introduce a class of macroscopic observables $N_\varphi(t)$ associated with a ``test'' function $\varphi$ which measure a specific property of the gas at time $t$, and formulate four theorems regarding the asymptotic limits of $N_\varphi(t)$ as $t\to \infty$, which establish phase space mixing for $d=1,2$ and different regularity assumptions on $\varphi$ and the initial one-particle distribution function. Also in section~\ref{Sec:Main} we introduce action-angle variables on phase space to obtain an explicit solution representation for the one-particle distribution function, provide an intuitive explanation for the mixing phenomenon, and discuss the case $d=3$. Complete proofs of our theorems, which exploit the action-angle variable representation and results from recent work by Mitchell~\cite{cM19} are given in Appendix~\ref{App:Proofs}. Next, in section~\ref{Sec:DarkMatterHalos} we apply our theorems to a Vlasov gas propagating in dark matter halos and other potentials relevant to stellar dynamics, providing an explanation for the virialization process observed numerically in Ref.~\cite{pDeJmAeMdN17}. In section~\ref{Sec:BlackHoles} we discuss mixing for a general relativistic gas configuration trapped in the potential well of a spherically symmetric, static black hole. Conclusions and a list of open questions for future work are given in section~\ref{Sec:Conclusions}. Technical, yet interesting and important issues are included in several appendices. Appendix~\ref{App:Proofs} contains complete proofs of the main theorems regarding the mixing property of one and two-dimensional systems formulated in section~\ref{Sec:Main}. Appendix~\ref{App:PeriodFunction} summarizes known results regarding the behavior of the period function $T(E)$ of bound orbits in one-dimensional systems, while Appendix~\ref{App:AreaFunction} lists analogous results for the behavior of the area function $A(E,L)$ corresponding to a bound orbit of energy $E$ and angular momentum $L$ in two-dimensional systems. Finally, Appendix~\ref{App:EffPotential} discusses the properties of the effective potential (relevant for the application in section~\ref{Sec:BlackHoles}) describing timelike geodesics  for a class of static, spherically symmetric and asymptotically flat black hole spacetimes.

%%%%%%%%%%%%%%%%%%%%%%%%%%%%%%%%%%%%%%%%%%%%%
\section{Newtonian model and mathematical results}
\label{Sec:Main}
%%%%%%%%%%%%%%%%%%%%%%%%%%%%%%%%%%%%%%%%%%%%%

In this section we formulate the relevant models for this article, which consist of a collisionless, simple, kinetic gas in $d$ space dimensions which is subject to an external potential $\Phi: U\to \Real$ defined on the configuration space $U\subset \Real^d$. For simplicity, we work in units for which the mass of each particle is one. The one-particle distribution function $f(t,x,p)$ which describes the state of the kinetic gas at time $t$ is a nonnegative function $f(t,\cdot,\cdot) : \Gamma\to \Real$ on phase space $\Gamma := U\times \Real^d$ which obeys the Vlasov equation\footnote{The Vlasov equation is also often called the collisionless Boltzmann or Liouville equation in the literature.}
\begin{equation}
\frac{\partial f}{\partial t} + \{ H , f \} =
\frac{\partial f}{\partial t} + p\cdot\nabla_x f - (\nabla_x\Phi)\cdot\nabla_p f = 0,
\label{Eq:Vlasov}
\end{equation}
with the Hamiltonian $H(x,p) = |p|^2/2 + \Phi(x)$. Given an initial distribution $f_0$ (in the space $L^1(\Gamma)$ of Lebesgue-integrable functions on $\Gamma$, say) describing the state of the gas at time $t = 0$, and under suitable assumptions on $\Phi: U\to \Real$ (see, for instance, section X.14 in Ref.~\cite{ReedSimon80}), Eq.~(\ref{Eq:Vlasov}) possesses a unique solution $f(t,\cdot,\cdot)$ for all $t\in \Real$ such that $f(0,\cdot,\cdot) = f_0$.

As explained in the introduction, macroscopic observables (including the particle number, energy, pressure and so on) are obtained from averaging the corresponding microscopic observable over phase space, where the distribution function enters as a weight function. More generally, let $\varphi: \Gamma\to \Real$ be a given test function (which, for the moment, we restrict to the space $C_0^\infty(\Gamma)$ of smooth functions of compact support), then we assign to it the corresponding (time-dependent) observable
\begin{equation}
N_\varphi(t) := \int\limits_\Gamma f(t,x,p) \varphi(x,p) d^d x d^d p,\qquad t\in\Real.
\label{Eq:Observable}
\end{equation}
Physically, the test function $\varphi$ can be thought of as a device which measures the properties of the system (such as its energy, mean velocity etc.) in a small region of phase space (corresponding to the support of $\varphi$), and $N_\varphi$ is the corresponding averaged value measured by this device.

The central problem addressed in this article is the asymptotic behavior of $N_\varphi(t)$ for $t\to \infty$, under the assumption that all the particle orbits are bounded in phase space. One may naively expect that in this case $N_\varphi(t)$ exhibits an everlasting oscillatory behavior. For example, for a one-dimensional gas of particles trapped in a potential well, the individual gas particles simply oscillate in time and (in the absence of dissipation) the motion on each energy surface is periodic. However, the key point is that, in general, different orbits have slightly different periods, such that an initially localized distribution function spreads in phase space, implying that the limit $\lim_{t\to\infty} N_\varphi(t)$ \emph{does} exist. Furthermore, as we will show for the simple linear model in Eq.~(\ref{Eq:Vlasov}) with a rotationally symmetric potential $\Phi$, this limit can even be predicted from the properties of the initial distribution function $f_0$ without performing a time evolution.

\subsection{One-dimensional model}

\begin{figure}[ht]
\centerline{\resizebox{9.0cm}{!}{\includegraphics{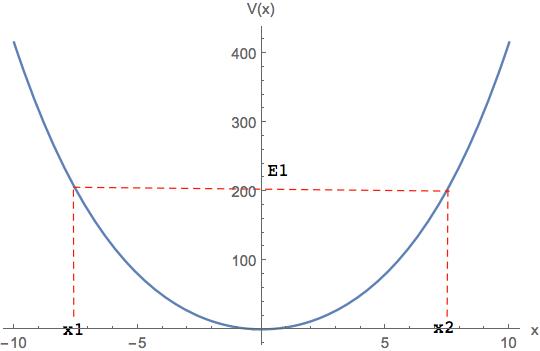}}\resizebox{7cm}{!}{\includegraphics{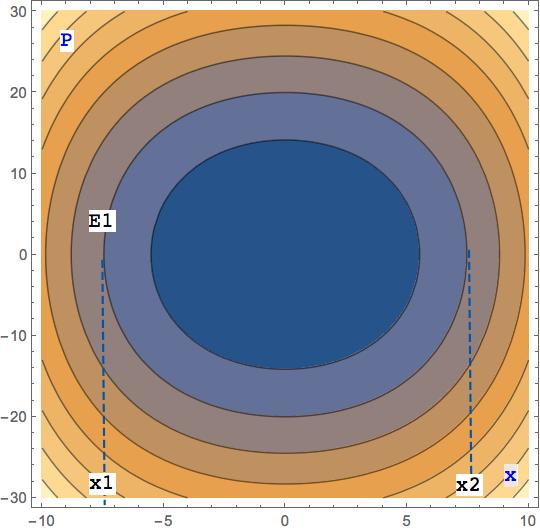}}}
\caption{\label{Fig:PotentialExample} A graph of the potential $V(x) = \frac{\omega_0}{2} x^2 + \frac{k}{4} x^4$, with parameter values $\omega_0 = 5.76$, $k = 0.051$ and the corresponding contour plot of the phase orbits. Also illustrated are the values of the turning points $x_1, x_2$ for the particular energy value $E_1 = 200$.}
\end{figure}

Our first model consists of a collisionless simple kinetic gas in one space dimension which is subject to an external potential $V: I\to \Real$ defined on an open interval $I = (a,b)$, $-\infty\leq a < b\leq \infty$ of $\Real$. (In the one-dimensional case we shall call the potential $V$ instead of $\Phi$.) We assume that $V\in C^\infty(I)$ is smooth and has non-vanishing derivative except at $x = x_0$ which is a non-degenerate global minimum of $V$, such that $V'(x) < 0$ for all $a < x < x_0$, $V'(x) > 0$ for all $x_0 < x < b$ and $V''(x_0) > 0$. Finally, we assume that $V(x)\to \infty$ when $x\to a$ or $x\to b$, implying that all the orbits with energy $E > E_0 := V(x_0)$ are bounded and periodic (see figure~\ref{Fig:PotentialExample})
with period
\begin{equation}
T(E) = 2\int\limits_{x_1(E)}^{x_2(E)} \frac{dx}{\sqrt{2(E - V(x))}},\qquad E > E_0.
\label{Eq:PeriodFunction}
\end{equation}
Here, $x_1(E) < x_2(E)$ denote the turning points of the orbit. As is well-known (see for instance~\cite{Arnold-Book}), the period function can be computed from the first derivative of the area function
\begin{equation}
A(E) := \oint\limits_{C(E)} p dx = 2\int\limits_{x_1(E)}^{x_2(E)} \sqrt{2(E - V(x))} dx,
\qquad E > E_0,
\label{Eq:AreaFunction1d}
\end{equation}
representing the area enclosed in phase space by the energy curve
\begin{equation}
C(E) := \left\{ (x,p)\in \Gamma : H(x,p) = \frac{p^2}{2} + V(x) = E \right\}.
\label{Eq:EnergyCurve}
\end{equation}

As will be discussed in section~\ref{SubSec:AA}, the solution of the Cauchy problem consisting of Eq.~(\ref{Eq:Vlasov}) with initial datum $f_0\in L^1(\Gamma)$ has a simple explicit representation in terms of action-angle variables; in this representation the evolution is given by a rotation by the angle $\omega t$ along each energy curve $C(E)$ with the ($E$-dependent) angular frequency $\omega = 2\pi/T(E)$. As such, the distribution function $f$ does not possess a pointwise limit when $t\to \infty$, since it is periodic on each energy surface. Nevertheless, as indicated before, one can prove that under a suitable non-degeneracy condition on $T(E)$ the macroscopic observable $N_\varphi(t)$ possesses a well-defined limit for any $\varphi\in C_0^\infty(\Gamma)$ as $t\to \infty$.  Moreover, this limit can be computed solely in terms of the initial datum. This is the content of the next theorem.

\begin{theorem}[One-dimensional phase space mixing]
\label{Thm:PSM1}
Let $f_0\in L^1(\Gamma)$ and let $f(t,x,p)$ be the unique solution of Eq.~(\ref{Eq:Vlasov}) with initial datum $f(0,x,p) = f_0(x,p)$ for all $(x,p)\in \Gamma$. Suppose further that for all $E > E_0$,
\begin{equation}
\frac{dT}{dE}(E) = \frac{d^2 A}{dE^2}(E) \neq 0.
\label{Eq:NonDegeneracyCond}
\end{equation}
Then, for all test functions $\varphi\in C^\infty_0(\Gamma)$ it holds that
\begin{equation}
\lim\limits_{t\to\infty} N_\varphi(t) 
 = \int\limits_\Gamma \langle f_0 \rangle_{C(E)}(x,p) \varphi(x,p) dx dp,
\label{Eq:RelaxationObs}
\end{equation}
where $\langle f_0 \rangle_{C(E)}$ denotes the average of $f_0$ over the energy curve $C(E)$ (see Eq.~(\ref{Eq:CEAverage}) below). Therefore, the macroscopic observable converges in time to the same observable obtained from averaging the initial distribution function $f_0$ over each orbit.
\end{theorem}

\proof See Appendix~\ref{App:Proofs}.
\qed

{\bf Remarks}:
\begin{enumerate}
\item One important physical consequence of the theorem is that as far as the macroscopic observables $N_\varphi$ are concerned, any initial gas configuration relaxes in time to a state that is described by the stationary distribution function $\langle f_0 \rangle_{C(E)}$ which depends only on energy $E$.
\item This relaxation process takes place even though the underlying Vlasov equation is reversible in time! The reason for this apparent paradox relies in the fact that even though at the level of the distribution function there is no (pointwise) convergence, the convergence takes place at the level of the observables, see the discussion in section 1.4 of~\cite{cMcV11} for further details.
\item It is possible to considerably weaken the hypothesis on the regularity and support of $\varphi$. In fact, it is sufficient to require $\varphi: \Gamma\to \Real$ to be continuous and bounded,  see Appendix~\ref{App:Proofs}.
\item The satisfaction of the non-degeneracy condition~(\ref{Eq:NonDegeneracyCond}) implies that the period function $T(E)$ must be monotonous. Sufficient conditions on the potential $V$ guaranteeing this property are derived in Refs.~\cite{sCdW86,cC87} and reviewed in Appendix~\ref{App:PeriodFunction}. However, as we will show in Appendix~\ref{App:Proofs}, the conclusion of the theorem still holds in case the non-degeneracy condition~(\ref{Eq:NonDegeneracyCond}) holds for almost all $E\in (E_0,\infty)$, that is, if it holds for all $E\in (E_0,\infty)\setminus Z$ with $Z$ a set of (Lebesgue) measure zero.
\item A relevant question is how fast the convergence~(\ref{Eq:RelaxationObs}) occurs and if there is an associated characteristic timescale. Unfortunately, this is not a simple question to answer since it depends on various factors (smoothness of $f_0$, support of the test function $\varphi$, magnitude of $dT/dE$), as can be understood from the proof in Appendix~\ref{App:Proofs}.
\item The conditions on the potential $V: I\to \Real$ can be relaxed. For example, if $V\in C^\infty(I)$ and $V$ has a non-degenerate local minimum at $x_0\in I$ with corresponding potential well $I_1 = (a_1,b_1)\subset I$, such that $V(a_1) = V(b_1)$, $V'(x) < 0$ for all $a_1 < x < x_0$, $V'(x) > 0$ for all $x_0 < x < b_1$ and $V''(x_0) > 0$, then the statement of the theorem still holds for any test function whose support lies inside the subset of $\Gamma$ corresponding to the well:
\begin{equation}
\Gamma_{well} = \{ (x,p) : x\in I_1, H(x,p) < V(a_1) = V(b_1) \}.
\end{equation}
\end{enumerate}

Besides the class of observables $N_\varphi(t)$ constructed from the distribution function $f$ via a smooth test function of compact support $\varphi\in C_0^\infty(\Gamma)$, one might also be interested in the particular quantiy
\begin{equation}
M_k(t,x_0) := \int\limits_{-\infty}^\infty f(t,x_0,p) p^k dp,
\qquad t > 0,\quad x_0\in I,
\label{Eq:MkDef}
\end{equation}
corresponding to the $k$'th moment of the distribution function relative to the momentum space. For example, $M_0(t,x_0)$ describes the particle density of the kinetic gas, $M_1(t,x_0)$ its linear momentum density, and $M_2(t,x_0)/2$ its kinetic energy density at the event $(t,x_0)$. Formally, the quantity $M_k(t,x_0)$ corresponds to the particular observables  $N_\varphi$ in Eq.~(\ref{Eq:Observable}) for which the test function is given by
\begin{equation}
\varphi(x,p) := p^k\delta(x - x_0).
\label{Eq:SpecialTestFunction}
\end{equation}
Since such test functions are distributional, Theorem~\ref{Thm:PSM1} cannot be applied. In fact, it is clear that the conclusion of the theorem cannot hold for an arbitrary distributional $\varphi$, as is already clear from the simple example $\varphi(x,p) = \delta(x - x_0)\delta(p - p_0)$ for which $N_\varphi(t) = f(t,x_0,p_0)$ does not converge for $t\to \infty$ in general. However, the next theorem shows that mixing still occurs for distributional test functions of the form~(\ref{Eq:SpecialTestFunction}), provided extra regularity of $f_0$ is required.

\begin{theorem}
\label{Thm:PSM2}
Let $f_0\in C^1_0(\Gamma)$ be continuously differentiable with compact support, and let $f(t,x,p)$ be the unique solution of Eq.~(\ref{Eq:Vlasov}) with initial datum $f(0,x,p) = f_0(x,p)$ for all $(x,p)\in \Gamma$. Suppose that the non-degeneracy condition~(\ref{Eq:NonDegeneracyCond}) is satisfied for almost all $E > E_0$.

Then, for all $x\in I$ and all $k\in \Natural_0$ it holds that
\begin{equation}
\lim\limits_{t\to\infty} M_k(t,x) 
 = \int\limits_{-\infty}^\infty \langle f_0 \rangle_{C(E)}(x,p) p^k dp.
\end{equation}
\end{theorem}

\proof See Appendix~\ref{App:Proofs}.
\qed

{\bf Remark}: As will be clear from the proof in Appendix~\ref{App:Proofs}, the statement of the theorem can be generalized to any test functions of the form $\varphi(x,p) = g(p)\delta(x - x_0)$ with a continuous function $g: \Real\to \Real$. The compact support hypothesis of $f_0$ can also be relaxed and replaced by a suitable fall-off condition on $f_0$ and its derivatives, see Appendix~\ref{App:Proofs}.

\subsection{Two-dimensional model}
\label{SubSec:2DModel}

Next, we consider a collisionless kinetic gas in two space dimensions which is subject to the external potential $\Phi: U\to \Real$ which we assume to be rotationally-invariant and defined on $U := \Real^2\setminus \{ (0,0) \}$. In terms of polar coordinates $(r,\varphi,p_r,p_\varphi)$ the Hamiltonian is
\begin{equation}
H(x,p) = \frac{1}{2}\left( p_r^2 + \frac{p_\varphi^2}{r^2} \right) + \Phi(r),
\end{equation}
where from now on, for simplicity, we regard $\Phi$ as a function of the single coordinate $r$. More precisely, we assume $\Phi: (0,\infty)\to \Real$ to be $C^\infty$-smooth, strictly monotonously increasing and such that the function $(0,\infty)\to (0,\infty)$, $r\mapsto r^3\Phi'(r)$ has positive derivative and increases monotonically from $0$ to $\infty$. These conditions imply for each $L\neq 0$ the existence of a unique global minimum of the effective potential
\begin{equation}
V_L(r) := \frac{L^2}{2r^2} + \Phi(r),\qquad r > 0,
\label{Eq:VLDef}
\end{equation}
whose location $r = r_0$ is determined by the equation $r_0^3\Phi'(r_0) = L^2$ and corresponds to a circular trajectory with minimum energy $E_0(L) = V_L(r_0)$. Note also that $V_L''(r_0) = r_0^{-3} \frac{d}{dr} [ r^3\Phi'(r) ]_{r=r_0} > 0$, such that the minimum is non-degenerate.

For the following, we restrict ourselves to the invariant subset
\begin{equation}
\Gamma_{bound} 
 := \{ (x,p)\in U\times \Real^2 : L = p_\varphi\neq 0, E_0(L)\leq H(x,p) < \Phi_\infty \}
\end{equation}
of phase space corresponding to non-radial bound trajectories, where $\Phi_\infty\in (-\infty,+\infty]$ denotes the asymptotic value of the potential.\footnote{Note that we exclude from $\Gamma_{bound}$ those trajectories which have vanishing angular momentum $L = 0$. This restriction can be dropped if $\Phi$ is regular at the center, in which case one sets $E_0(0)  = \Phi(0)$, but otherwise $L=0$ should be excluded since the corresponding trajectories are incomplete.} For the statement of the following theorem the area function $A : \Omega\to \Real$ defined by
\begin{eqnarray}
&& \Omega := \{ (E,L)\in \Real^2 : L\neq 0, E_0(L) < E < \Phi_\infty \},
\label{Eq:AreaOmega}\\
&& A(E,L) := \oint p_r dr = 2\int\limits_{r_1(E,L)}^{r_2(E,L)} \sqrt{2(E - V_L(r))} dr,\qquad
(E,L)\in \Omega,
\label{Eq:AreaFunction}
\end{eqnarray}
with turning points $r_1(E,L) < r_2(E,L)$ plays an important role.

\begin{theorem}[Two-dimensional phase space mixing in rotationally symmetric potential]
\label{Thm:PSM3}
\mbox{Let $f_0\in L^1(\Gamma_{bound})$} and let $f(t,x,p)$ be the unique solution of Eq.~(\ref{Eq:Vlasov}) with initial datum $f(0,x,p) = f_0(x,p)$ for all $(x,p)\in \Gamma_{bound}$. Suppose further that the condition
\begin{equation}
\det(D^2 A(E,L)) \neq 0,
\label{Eq:NonDegeneracyCond2D}
\end{equation}
holds for almost all $(E,L)\in \Omega$, where $D^2 A(E,L)$ denotes the Hessian matrix of the area function~(\ref{Eq:AreaFunction}).

Then, for all test functions $\varphi\in C^\infty_0(\Gamma_{bound})$ it holds that
\begin{equation}
\lim\limits_{t\to\infty} N_\varphi(t) 
 = \int\limits_{\Gamma_{bound}} \langle f_0 \rangle(x,p) \varphi(x,p) d^2x d^2p,
\end{equation}
where $\langle f_0 \rangle$ denotes the average of $f_0$ over the angle variables (see Eq.~(\ref{Eq:AngleAverage}) below).
\end{theorem}

\proof See Appendix~\ref{App:Proofs}.
\qed

{\bf Remarks}:
\begin{enumerate}
\item Condition~(\ref{Eq:NonDegeneracyCond2D}) is the two-dimensional generalization of the non-degeneracy condition~(\ref{Eq:NonDegeneracyCond}). More details on the satisfaction of this condition for specific examples are discussed in section~\ref{Sec:DarkMatterHalos} and Appendix~\ref{App:AreaFunction}.
\item As in the one-dimensional case, the conditions on $\varphi$ can be weakened: it is in fact sufficient to requiere $\varphi: \Gamma_{bound}\to \Real$ to be a continuous and bounded function.
\item It is interesting to ask what happens in the complementary region $\Gamma_{unbounded} = \Gamma\setminus \Gamma_{bound}$ of phase space corresponding to unbounded trajectories. Assuming that $L\neq 0$, such trajectories have the property of bouncing off the centrifugal potential; hence one expects that for $f\in L^1(\Gamma)$ and $\varphi\in C^\infty_0(\Gamma_{unbound})$ one should find
\begin{equation}
\lim\limits_{t\to \infty} N_\varphi(t) = 0.
\end{equation}
Although in this article we leave the proof of this statement open, we point out that  a related result has been established in the general relativistic setting of a collisionless kinetic gas accreted by a Schwarzschild black hole, see section~5.2 in Ref.~\cite{pRoS16}.
\end{enumerate}

As in the one-dimensional case, one can show that (under extra regularity conditions on $f_0$) the conclusion of the theorem still holds for certain distributional test functions. Specifically, let $g:\Real^2\to \Real$ be continuous, and consider for $x_0\in U$ the test function
\begin{equation}
\varphi(x,p) = \delta(x - x_0) g(p).
\end{equation}
Then, one can show:

\begin{theorem}
\label{Thm:PSM4}
Let $f_0\in C^2_0(\Gamma_{bound})$ be twice continuously differentiable with compact support, and let $f(t,x,p)$ be the unique solution of Eq.~(\ref{Eq:Vlasov}) with initial datum $f(0,x,p) = f_0(x,p)$ for all $(x,p)\in \Gamma_{bound}$. Suppose further that the condition~(\ref{Eq:NonDegeneracyCond2D}) holds for almost all $(E,L)\in \Omega$.

Then,
\begin{equation}
\lim\limits_{t\to\infty} \int\limits_{H(x_0,p) < \Phi_\infty} f(t,x_0,p) g(p) d^2 p 
 = \int\limits_{H(x_0,p) < \Phi_\infty} \langle f_0 \rangle(x_0,p) g(p) d^2p
\end{equation}
for all continuous functions $g: \Real^2\to \Real$.
\end{theorem}

\proof See Appendix~\ref{App:Proofs}.
\qed

In the following subsection, we reformulate the one and two-dimensional problems in terms of action-angle variables which provide an explicit solution representation for the one-particle distribution function and provides some intuitive ideas about the mixing phenomenon. Once this has been achieved, the proofs of Theorems~\ref{Thm:PSM1} and \ref{Thm:PSM3} can be established using the results in Ref.~\cite{cM19} which, for completeness, are summarized in Appendix~\ref{App:Proofs}. The proofs of Theorems~\ref{Thm:PSM2} and \ref{Thm:PSM4}, for which the test function is distributional, are also given in Appendix~\ref{App:Proofs}, and they are based on an extension of the results in~\cite{cM19}.

\subsection{Action-angle variables and explicit solution representation for the distribution function}
\label{SubSec:AA}

We start with the one-dimensional problem ($d=1$) and introduce action-angle variables $(Q,J)$ on phase space $\Gamma$.\footnote{See, for instance, Ref.~\cite{Arnold-Book} for an introduction to action-angle variables and Theorem~V.40 in~\cite{Zehnder-Book} for a precise statement.} For this, we first recall the definition of the area function $A(E)$ in Eq.~(\ref{Eq:AreaFunction1d}) which satisfies $dA/dE = T > 0$ and $A(E)\to 0$ for $E\to E_0$ while $A(E)\to \infty$ as $E\to \infty$.\footnote{A rough estimate yields, for any $E > E^* > E_0$,
$$
A(E) \geq 2\int\limits_{x_1^*}^{x_2^*} \sqrt{2(E - E^*)} dx = 2(x_2^* - x_1^*)\sqrt{2(E - E^*)},\qquad
x_j^* := x_j(E^*),
$$
which shows that $A(E)\to \infty$ as $E\to \infty$.} Therefore, $A: (E_0,\infty)\to (0,\infty)$ is a smooth, invertible function. The action variable $J$ is defined by
\begin{equation}
J(x,p) := \frac{A(H(x,p))}{2\pi} = \frac{1}{2\pi}\oint\limits_{C(E)} p dx.
\label{Eq:JDef}
\end{equation}
By definition, $J$ is constant along each energy surface $C(E)$ and since $A$ is monotonous, $J$ provides a unique label for each energy surface $C(E)$. The corresponding angle variable is defined by
\begin{equation}
Q(x,p) := \frac{2\pi}{T(H(x,p))}\int\limits_{\gamma_x} \frac{dx}{p},
\label{Eq:QDef}
\end{equation}
where $\gamma_x$ is a curve in $\Gamma$, oriented clock-wise, which connects the left turning point $x_1(E)$ with the given point $(x,p)$ along the curve $C(E)$ with energy $E = H(x,p)$. A full revolution (in the clock-wise direction) along $C(E)$ induces the change $Q\mapsto Q + 2\pi$; hence the variable $Q$ provides an angle along each energy curve $C(E)$. It is well-defined everywhere on phase space $\Gamma$ except at the critical point $(x,p) = (x_0,0)$ of the Hamiltonian.

By construction (and since $H$ is constant along $\gamma_x$), one has $dJ\wedge dQ = dp\wedge dx$. Hence, the pair $(Q,J)$ defines smooth symplectic coordinates on
\begin{equation}
\Gamma_0 := \Gamma\setminus \{ (x_0,0) \}\simeq S^1\times (0,\infty),
\end{equation}
the subset of phase space obtained by omitting the equilibrium point. In terms of the new coordinates $(Q,J)$ the Hamiltonian becomes a function of $J$ only, and hence the Vlasov equation~(\ref{Eq:Vlasov}) reduces to
\begin{equation}
\frac{\partial f}{\partial t} + \omega(J)\frac{\partial f}{\partial Q} = 0,\qquad
\omega(J) := \left. \frac{\partial H}{\partial J} \right|_Q
 = \left( \frac{1}{2\pi}\frac{dA}{dE}(E) \right)^{-1} = \frac{2\pi}{T(E)},
\label{Eq:VlasovAA1D}
\end{equation}
whose solution is
\begin{equation}
f(t,x,p) = F(Q - \omega(J)t, J),
\end{equation}
where the function $F: S^1\times (0,\infty)\to \Real$ is determined by the initial datum, i.e. $F(Q,J) = f_0(x,p)$ for all $(x,p)\in \Gamma_0$. It follows from Eq.~(\ref{Eq:VlasovAA1D}) that the time evolution of the distribution function consists of a rotation along each energy surface $C(E)$ with constant angular velocity $\omega(J)$. In the completely degenerate case where $\omega$ is independent of the energy surface $C(E)$, each of these rotations is in phase and there is no mixing in phase space. However, when the frequencies $\omega(J)$ are different from each other, the phase flow stretches and spreads the distribution function over the phase space, giving rise to the mixing property. Note that the condition $d\omega/dJ \neq 0$ is equivalent to $dT/dE \neq 0$. As we show in Appendix~\ref{App:Proofs}, at the level of the observables, this effect allows one to replace the distribution function $f$ with its average over $C(E)$:
\begin{equation}
\langle f \rangle_{C(E)} = \langle f_0 \rangle_{C(E)} 
 := \frac{1}{2\pi}\int\limits_0^{2\pi} F(Q,J) dQ.
\label{Eq:CEAverage}
\end{equation}

Next, let us describe the situation in $d=2$ dimensions. On the phase space $\Gamma_{bound}$ of bound trajectories, the action-angle variables $(Q_r,Q_\varphi,J_r,J_\varphi)$ are defined as follows. Denote by
\begin{equation}
\Gamma_{E,L} := \{ (x,p)\in \Gamma_{bound} : H(x,p) = E, p_\varphi = L \}
\end{equation}
the invariant submanifold corresponding to the integrals of motion $(E,L)\in \Omega$, where $\Omega$ is defined in Eq.~(\ref{Eq:AreaOmega}). Each of these submanifolds is topologically a two-torus $T^2$. Let $C_r(E,L)$ be the closed loop in the $(r,p_r)$ plane around $\Gamma_{E,L}$, and likewise $C_\varphi(E,L)$ the closed loop around the azimuthal direction $\varphi$. Then, we define
\begin{equation}
I_r(E,L) := \frac{1}{2\pi}\oint\limits_{C_r(E,L)} p_r dr  = \frac{A(E,L)}{2\pi},\qquad
I_\varphi(E,L) :=  \frac{1}{2\pi}\oint\limits_{C_\varphi(E,L)} p_\varphi d\varphi = L,
\label{Eq:IMap}
\end{equation}
where $A(E,L)$ is the area function defined in Eq.~(\ref{Eq:AreaFunction}). This provides a smooth transformation $I: \Omega\to (0,\infty)\times\Real\setminus \{ 0 \}$, $(E,L)\mapsto (I_r(E,L),I_\varphi(E,L)) = (A(E,L)/2\pi,L)$ whose Jacobi matrix is
\begin{equation}
DI(E,L) = \left(\begin{array}{cc}
\frac{T(E,L)}{2\pi} & -\frac{\Delta\psi(E,L)}{2\pi} \\
0 & 1 
\end{array} \right),
\label{Eq:JacobiMatrix}
\end{equation}
with
\begin{eqnarray}
T(E,L) &:=& \frac{\partial A}{\partial E}(E,L) 
= 2\int\limits_{r_1(E,L)}^{r_2(E,L)} \frac{dr}{\sqrt{2(E - V_L(r))}} > 0,
\label{Eq:TDef}\\
\Delta\psi(E,L) &:=& -\frac{\partial A}{\partial L}(E,L) 
 = 2L\int\limits_{r_1(E,L)}^{r_2(E,L)} \frac{1}{\sqrt{2(E - V_L(r))}}\frac{dr}{r^2},
\label{Eq:DPsiDef}
\end{eqnarray}
describing, respectively, the period of the orbit in the $(r,p_r)$- plane and the azimuthal shift during one such period (and hence the trajectory is closed if and only if $\Delta\psi$ is a rational multiple of $2\pi$). For the following, we denote by $\Omega_J := I(\Omega)$ the image of $I$. Then, it is not difficult to see that $\Omega_J$ is open and that $I: \Omega\to \Omega_J$ is invertible since for each fixed $L\neq 0$ the period function $T(E,L)$ is positive, implying that the area function $A(E,L)$ is monotonously increasing in $E$.

The action variables are defined as:
\begin{equation}
J_r(x,p) := I_r(H(x,p),p_\varphi),\qquad
J_\varphi(x,p) := I_\varphi(H(x,p),p_\varphi),
\end{equation}
while the angle variables are defined by
\begin{equation}
Q_r(x,p) := \frac{\partial S}{\partial J_r}(x; J(x,p)),\qquad
Q_\varphi(x,p) := \frac{\partial S}{\partial J_\varphi}(x; J(x,p)),
\end{equation}
with the generating function
\begin{equation}
S(x; I) := I_\varphi\varphi + \int\limits_{\gamma_r} p_r dr,
\end{equation}
with $\gamma_r$ a curve along $C_r(E,L)$ which connects the left turning point $r_1(E,L)$ with the given point $r$. Explicit calculation using the inverse transposed of the matrix~(\ref{Eq:JacobiMatrix}) gives
\begin{eqnarray}
Q_r(x,p) &=& \left. \frac{2\pi}{T(E,L)} \int\limits_{\gamma_r} \frac{dr}{p_r} \right|_{(E,L) = (H(x,p),p_\varphi)},
\\
Q_\varphi(x,p) &=& \left[ \varphi - L\int\limits_{\gamma_r} \frac{1}{p_r}\frac{dr}{r^2}
 + \frac{\Delta\psi(E,L)}{T(E,L)}\int\limits_{\gamma_r} \frac{dr}{p_r} \right]_{(E,L) = (H(x,p),p_\varphi)},
\end{eqnarray}
where the denominator $p_r$ in the integrals is determined by the energy equation $p_r^2/2 + V_L(r) = E$. Note that along a full revolution around $C_r(E,L)$ one has $(Q_r,Q_\varphi)\mapsto (Q_r + 2\pi,Q_\varphi)$ while along a full revolution around $C_\varphi(E,L)$ one has $(Q_r,Q_\varphi)\mapsto (Q_r,Q_\varphi + 2\pi)$, such that $(Q_r,Q_\varphi)$ define angles parametrizing the tori $\Gamma_{E,L}$. In this way, one obtains smooth symplectic coordinates on
\begin{equation}
\Gamma_0 := \bigcup\limits_{(E,L)\in \Omega} \Gamma_{E,L}
 = \Gamma_{bound}\setminus \Gamma_{circ}, 
\end{equation}
i.e. the phase space consisting of bound trajectories minus the circular trajectories $\Gamma_{circ} = \{ (x,p) : H(x,p) = E_0(L), L = p_\varphi \}$.

Expressing the Hamiltonian $H$ in terms of the action-angle variables $(J_r,J_\varphi,Q_r,Q_\varphi)$, the Vlasov equation~(\ref{Eq:Vlasov}) reads
\begin{equation}
\frac{\partial f}{\partial t} + \omega_r(J_r,J_\varphi)\frac{\partial f}{\partial Q_r} 
 + \omega_\varphi(J_r,J_\varphi)\frac{\partial f}{\partial Q_\varphi} = 0,
\label{Eq:VlasovAA}
\end{equation}
with the fundamental frequencies $\omega_r$ and $\omega_\varphi$ associated with the motion in the radial and azimuthal directions given by
\begin{eqnarray}
\omega_r(J_r,J_\varphi) &:=& \frac{\partial H}{\partial J_r} 
 = \left. \frac{2\pi}{T(E,L)} \right|_{(E,L) = I^{-1}(J)},
\label{Eq:omega_r}\\
\omega_\varphi(J_r,J_\varphi) &:=& \frac{\partial H}{\partial J_\varphi}
 = \left. \frac{\Delta\psi(E,L)}{T(E,L)} \right|_{(E,L) = I^{-1}(J)}.
\label{Eq:omega_phi}
\end{eqnarray}
The general solution of Eq.~(\ref{Eq:VlasovAA})  has the form
\begin{equation}
f(t,x,p) = F(Q_r - \omega_r(J_r,J_\varphi)t,Q_\varphi - \omega_\varphi(J_r,J_\varphi)t, J_r,J_\varphi),
\end{equation}
with the function $F: T^2\times\Omega_J\to \Real$ determined by the initial datum: $F(Q_r,Q_\varphi,J_r,J_\varphi) = f_0(x,p)$ for all $(x,p)\in \Gamma_0$. If the frequencies $\omega_r$ and $\omega_\varphi$ are constant, all the particles wind around the invariant two-tori $T^2$ in phase and there is no mixing. In contrast to this, and as proven in Appendix~\ref{App:Proofs}, mixing takes place such that the distribution function can be replaced with its angle-average
\begin{equation}
\langle F \rangle(Q,J) := \frac{1}{(2\pi)^2}\int\limits_{T^2} F(\overline{Q},J) d^2\overline{Q},
\label{Eq:AngleAverage}
\end{equation}
if the frequency map $\omega: \Omega_J\to \Real^2$, $(J_r,J_\varphi)\mapsto (\omega_r,\omega_\varphi)$ satisfies the condition
\begin{equation}
\det( D\omega(J_r,J_\varphi) ) = \det( D^2 H(J_r,J_\varphi) ) = \det\left( \begin{array}{cc}
\frac{\partial \omega_r}{\partial J_r} & \frac{\partial \omega_r}{\partial J_\varphi} \\
\frac{\partial \omega_\varphi}{\partial J_r} & \frac{\partial \omega_\varphi}{\partial J_\varphi}
\end{array} \right) \neq 0,
\label{Eq:HessianH}
\end{equation}
for almost all $(J_r,J_\varphi) \in \Omega_J$. Using Eqs.~(\ref{Eq:omega_r},\ref{Eq:omega_phi}) and the Jacobi matrix~(\ref{Eq:JacobiMatrix}) one finds
\begin{equation}
\det( D^2 H(J_r,J_\varphi) ) = \frac{(2\pi)^2}{T(E,L)^4}\det( D^2 A(E,L) ),
\end{equation}
and hence the condition~(\ref{Eq:HessianH}) is equivalent to the hypothesis~(\ref{Eq:NonDegeneracyCond2D}) in Theorem~\ref{Thm:PSM3}.

\subsection{Comments regarding the three-dimensional case}
\label{SubSec:3DModel}

The setup described in section~\ref{SubSec:2DModel} can easily be generalized to $d = 3$ or higher-dimensions, including the definitions of $\Gamma_{bound}$ and the area function $A(E,L)$. However, in this case the conclusions of Theorems~\ref{Thm:PSM3} and~\ref{Thm:PSM4} cannot be generalized in a straightforward way. To explain the reason for this, let us focus on the three-dimensional case and go back to the construction of action-angle variables $(Q_r,Q_\vartheta,Q_\varphi,J_r,J_\vartheta,J_\varphi)$ for $ d = 3$. The generalization of Eq.~(\ref{Eq:IMap}) is (see, for instance section 3.5.2 in Ref.~\cite{BinneyTremaine-Book})
\begin{equation}
I_r(E,L,L_z) = \frac{A(E,L)}{2\pi},\qquad
I_\vartheta(E,L,L_z) = L - |L_z|,\qquad
I_\varphi(E,L,L_z) = L_z,
\label{Eq:IMap3d}
\end{equation}
where now $L$ refers to the total angular momentum and $L_z$ to the azimuthal one, and $0 < |L_z| < L$. Instead of $(I_r,I_\vartheta,I_\varphi)$ it is simpler to consider the functions $I := (I_1,I_2,I_3) := (I_r,I_\vartheta + |I_\varphi|,I_\varphi) = (A(E,L)/2\pi,L,L_z)$ and the corresponding action variables $(J_1,J_2,J_3)$, which are obtained from these by the replacements
\begin{equation}
E\mapsto H(x,p),\quad
L\mapsto \sqrt{p_\vartheta^2 + \frac{p_\varphi^2}{\sin^2\vartheta}},\quad
L_z\mapsto p_\varphi.
\end{equation}
The Jacobi matrix of the map $I$ is
\begin{equation}
DI(E,L,L_z) = \left(\begin{array}{ccc}
\frac{T(E,L)}{2\pi} & -\frac{\Delta\psi(E,L)}{2\pi} & 0\\
0 & 1  & 0 \\
0 & 0 & 1
\end{array} \right),
\label{Eq:JacobiMatrix3d}
\end{equation}
with $T(E,L)$ and $\Delta\psi(E,L)$ defined as in Eqs.~(\ref{Eq:TDef},\ref{Eq:DPsiDef}). Since the Hamiltonian is a function of $J_1$ and $J_2$ alone, the fundamental frequencies are
\begin{eqnarray}
&& \omega_1(J) = \frac{\partial H}{\partial J_1}(J)
 = \left. \frac{2\pi}{T(E,L)} \right|_{(E,L,L_z) = I^{-1}(J)},\qquad
\omega_2(J) = \frac{\partial H}{\partial J_2}(J) 
 = \left. \frac{\Delta\psi(E,L)}{T(E,L)} \right|_{(E,L,L_z) = I^{-1}(J)},\\
&& \omega_3(J) = \frac{\partial H}{\partial J_3}(J) = 0.
\end{eqnarray}
Since $\omega_3$ is identically zero, the Jacobian of the frequency function is degenerated, and hence one cannot replace $F$ by its average over all $Q$'s. However, the results in the previous subsections imply that mixing still occurs in the angle variables $Q^1$ and $Q^2$, provided the area function $A(E,L)$ satisfies the non-degeneracy condition~(\ref{Eq:NonDegeneracyCond2D}). In this case, the observables converge in time to a state described by averaging the initial distribution function over $(Q^1,Q^2)$. The constant angle $Q^3$ on which this average still depends describes the angle between  the $x$-axis and the intersection of the orbital plane with the $xy$-plane, see section 3.5.2 in Ref.~\cite{BinneyTremaine-Book}. Therefore, mixing occurs within each orbital plane with the final state depending only on the integrals of motion (namely, the energy and the three components $(L_x,L_y,L_z)$ of the angular momentum) provided the non-degeneracy condition~(\ref{Eq:NonDegeneracyCond2D}) holds.

%%%%%%%%%%%%%%%%%%%%%%%%%%%%%%%%%%%%%%%%%%%%%
\section{Application to stellar systems and dark matter halos}
\label{Sec:DarkMatterHalos}
%%%%%%%%%%%%%%%%%%%%%%%%%%%%%%%%%%%%%%%%%%%%%

After discussing the mathematical results describing the mixing phenomenon for a collisionless gas in an exterior Newtonian central potential, in this section and the next one we provide a few  applications to astrophysical systems. In this section, we consider five examples of central potentials $\Phi(r)$ which arise in simple models of stellar systems or dark matter halos, and study the occurrence of mixing in $d=1$ and $d=2$ dimensions for each of these potentials.

Table~\ref{Tab:Potentials} summarizes these potentials, their asymptotic values and their corresponding mass density $\rho$. The first one is the isochrone potential (see~\cite{BinneyTremaine-Book} and references therein), which for large distances behaves like Kepler's potential, but is regular at the center $r = 0$ and corresponds to a localized smooth density distribution (falling off like $1/r^4$ for $r \gg 1 $). An interesting property of the isochrone potential is that all orbits can be determined analytically (see section 3.5.2 in~\cite{BinneyTremaine-Book}). The remaining four potentials arise in the modeling of dark matter halos, and their inclusion in our list has been motivated by the work in Ref.~\cite{pDeJmAeMdN17} which numerically solves the Vlasov equation for a spherically symmetric configuration with fixed angular momentum $L$. The first of these potentials is the isothermal potential, whose circular orbits have a constant velocity profile, and hence serves as a simple model to reproduce the velocity profile in the exterior region of galaxies. However, the associated density diverges at the center; to avoid this behavior one can replace it with the truncated isothermal potential which is regular at $r = 0$ and preserves the asymptotic form of the isothermal potential for $r\gg 1$. On the other hand, the Navarro-Frenk-White (NFW) potential has been obtained from the study of N-body simulations~\cite{jNcFsW96,jNcFsW97}, providing a ``universal" density profile for cold dark matter halos. This profile is shallower than the isothermal $1/r^2$ density near the center but steeper than $1/r^2$ in its outer region. The Burkert model~\cite{aB95} has the same fitting qualities than the NFW one, but has a finite density at the center (see~\cite{pDeJmAeMdN17} for more details and references). Note that all these potentials except the isochrone one have a mass function $M(r)$ (describing the mass enclosed in a sphere of radius $r$) which diverges at $r\to \infty$, indicating that they do not describe isolated systems.

\begin{table}[H]
    \centering
       \begin{tabular}{p{2cm}p{4.5cm}p{8.5cm}p{2.5cm}} 
         \hline \hline
      \textbf{name}&\textbf{density $\rho(r)$} & \textbf{potential $\Phi(r)$} & \textbf{asymptotic value $\Phi_\infty$}  \\
        \hline \hline 
    isochrone & $\frac{3+2r^2+3\sqrt{1+r^2}}{3\left(1+r^2 \right)^{3/2} \left(1+\sqrt{1+r^2} \right)^3 } $ & $ -\frac{1}{1+\sqrt{1+r^2}}$ & 0  \\     \hline \\
     isothermal              & $\frac{1}{r^2}$ & $3\log(r)$ & $\infty$ \\    \hline \\
     truncated isothermal & $\frac{10}{9}\frac{1}{1 + r^2}$ & $\frac{5}{3}\left( \log(1+r^2) + 2 \frac{\arctan (r)}{r} \right)$& $\infty$ \\   \hline \\      
     NFW & $\frac{16}{3r}\frac{1}{(1+r)^2}$ & $-16\frac{\log(1+r)}{r}$ & 0 \\ \hline \\
     Burkert & $\frac{40}{9}\frac{1}{(1+r)(1+r^2)}$ &
      $\frac{10}{3r} \left\{ 2(1+r) \left[  \arctan(r)-\log(1+r) \right] -(1-r) \log(1+r^2) \right\}$ & $\frac{10\pi}{3}$ \\  
       \hline 
        \end{tabular}
    \caption{\label{Tab:Potentials} List of (dimensionless) density profiles and corresponding Newtonian potentials analyzed in this section. Here, the physical density $\rho_{phys}$, potential $\Phi_{phys}$ and radius $r_{phys}$ are given by $\rho_{phys}(r_{phys}) = \rho_0\rho(r)$, $\Phi_{phys}(r_{phys}) = (G M_0/R_0) \Phi(r)$ and $r_{phys} = R_0 r$, with $\rho_0$, $M_0$ and $R_0$ a characteristic density, mass and radius which are related to each other through the relation $M_0 = 4\pi\rho_0 R_0^3/3$. As in~\cite{pDeJmAeMdN17} (which uses the notation $d$ for the dimensionless radius $r$), the last four potentials have been scaled such that their densities have the same value at $r = 3$.}
\end{table}

One can easily verify that the conditions on the potential $\Phi$ listed at the beginning of section~\ref{SubSec:2DModel} are satisfied for all the examples in Table~\ref{Tab:Potentials}, such that the effective potential $V_L(r)$ defined in Eq.~(\ref{Eq:VLDef}) has for each $L^2 > 0$ a unique global minimum describing circular trajectories.

\subsection{Mixing of a spherically symmetric distribution with fixed $L$}

In Ref.~\cite{pDeJmAeMdN17} the Vlasov equation~(\ref{Eq:Vlasov}) is solved numerically for a spherically symmetric scenario described by a one-particle distribution function proportional to $\delta(L - L_0)$, corresponding to a situation in which all particles have the same total angular momentum $L_0$. This reduces the problem to an effective one-dimensional problem described by the effective potential in Eq.~(\ref{Eq:VLDef}) for fixed $L = L_0$, for which Theorems~\ref{Thm:PSM1} and \ref{Thm:PSM2} apply. The main motivation in~\cite{pDeJmAeMdN17} was to analyze the propagation of small matter inhomogeneities in the dark matter halos given in Table~\ref{Tab:Potentials}, described by a given initial localized distribution function $f_0$ (the self-gravity generated by the kinetic gas is neglected, such that the potential $\Phi$ remains unchanged throughout the evolution). The results of the simulation show that after a transient period the distribution function reaches a ``stationary state", satisfying the virial theorem. In particular, we draw the reader's attention to figures~8--11 in~\cite{pDeJmAeMdN17} which show snapshots of the distribution function at different times. These figures clearly illustrate the mixing phenomenon taking place in phase space.

\begin{figure}[ht]
\centerline{\resizebox{9.0cm}{!}{\includegraphics{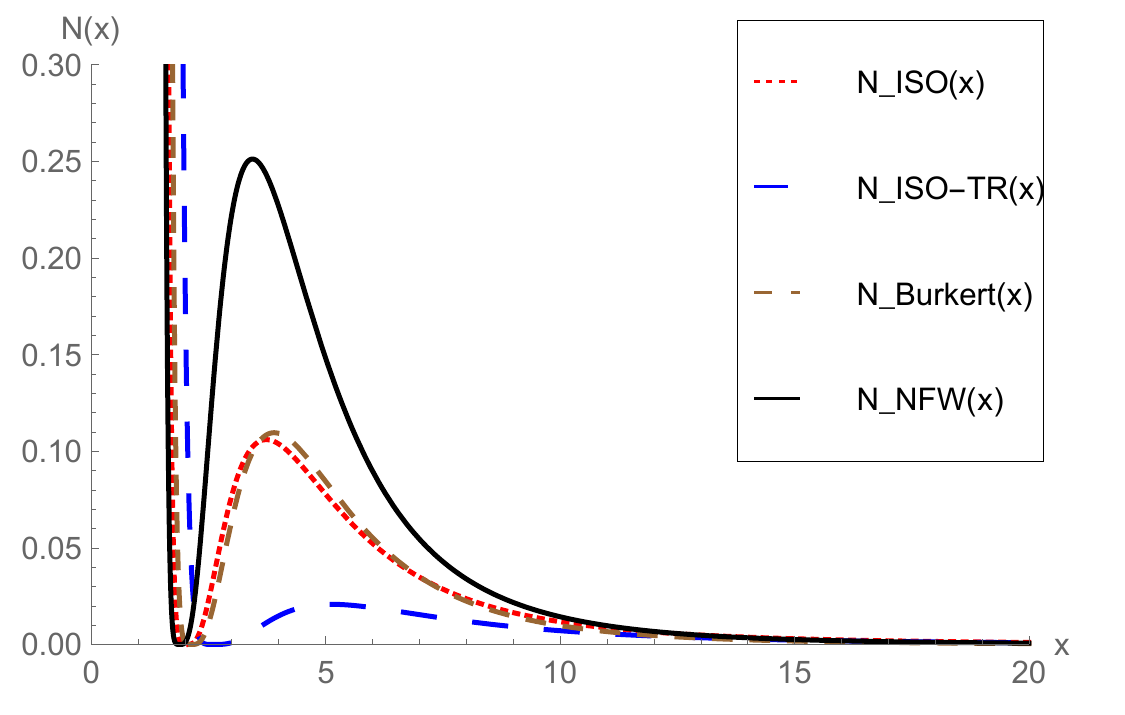}}}
\caption{\label{Fig:N(x)} A plot of the function $N(x)$ defined in Eq.~(\ref{Eq:NDef}), defined for the effective potential $V_L$ with $L = L_0 = 3.5$ belonging to each of the four dark matter potentials. As is visible from the plots (and proven rigorously for the isothermal case in Appendix~\ref{App:PeriodFunction}), the function $N(x)$ is positive in each case, implying that the period function $T(E)$ is strictly increasing.}
\end{figure}

More precisely, the simulations in figures 8--11 of Ref.~\cite{pDeJmAeMdN17} are based on Gaussian initial data of the form
\begin{eqnarray}
f_0(r,p) = \frac{N_0}{8\pi^3 L_0\sigma_d \sigma_p }
\left( e^{-\frac{(r - r_0)^2}{{\sigma_d}^2} - \frac{(p - p_0)^2}{{\sigma_p}^2}}
 + e^{-\frac{(r + r_0)^2}{{\sigma_d}^2} - \frac{(p + p_0)^2}{{\sigma_p}^2}} \right),
\label{Eq:InitialDistribution}
\end{eqnarray}
with parameter values $N_0 = 1$, $(r_0,p_0) = (3,-0.5)$, $(\sigma_d,\sigma_p) = (0.5,0.25)$, and $L_0 = 3.5$ for the angular momentum. According to Theorems~\ref{Thm:PSM1} and \ref{Thm:PSM2} (see also remark~6 following Theorem~\ref{Thm:PSM1}), mixing takes place provided the non-degeneracy condition~(\ref{Eq:NonDegeneracyCond}) is satisfied for (almost) all $E\in (E_0,\Phi_\infty)$, with $E_0$ denoting the minimal energy and $\Phi_\infty$ the asymptotic value of the potential. Using the methods described in Appendix~\ref{App:PeriodFunction} to analyze the sign of the derivative of the period function $T(E)$, we have verified that in fact the period function $T(E)$ is a strictly increasing function of $E$ for the four dark matter potentials, such that condition~(\ref{Eq:NonDegeneracyCond}) is satisfied in all cases. Figure~\ref{Fig:N(x)} shows for each of these potentials the function $N(x)$ defined in Eq.~(\ref{Eq:NDef}) (where here $x = r$), and the fact that it is positive implies the aforementioned monotonicity of $T(E)$.

Since mixing takes place, the gas configuration converges in time to a configuration that is described (in what the observables are concerned) by the averaged distribution function $\langle f_0 \rangle_{C(E)}$, obtained from the initial distribution function~(\ref{Eq:InitialDistribution}) by averaging over the energy curves $C(E)$. Since along $C(E)$ one has $dQ = (2\pi/T(E)) dx/p$, Eq.~(\ref{Eq:CEAverage}) for this average yields
\begin{equation}
\langle f_0 \rangle_{C(E)} 
 = \frac{1}{T(E)}\oint\limits_{C(E)} f_0(r,p) \frac{dr}{p}
 = \frac{1}{T(E)}\left. \int\limits_{r_1(E)}^{r_2(E)} \left[ f_0(r,p) + f_0(r,-p) \right]
 \frac{dr}{p} \right|_{p = \sqrt{2(E - V_L(r))}}.
 \label{Eq:f0Average1}
 \end{equation}
Alternatively, one can use the representation~(\ref{Eq:TE}) from Appendix~\ref{App:PeriodFunction} and compute this average by means of the formula:
 \begin{equation}
 \langle f_0 \rangle_{C(E)} 
 = \left. \frac{\int\limits_{0}^{2\pi} f_0(H(\varrho\sin\alpha),\varrho\cos\alpha) H'(\varrho\sin\alpha) d\alpha}
 {\int\limits_{0}^{2\pi} H'(\varrho\sin\alpha) d\alpha} \right|_{\varrho = \sqrt{2(E- E_0)}},
\label{Eq:f0Average2}
\end{equation}
which requires the computation of the function $H$ defined below Eq.~(\ref{Eq:hDef}), but has the advantage over the previous formula of involving integrands which are regular and do not diverge at the turning points. We have in fact computed the average $\langle f_0 \rangle_{C(E)}$ numerically using both formulae~(\ref{Eq:f0Average1},\ref{Eq:f0Average2}) and checked that the results agree with each other within a relative error of $0.5 \%$ in the maximum norm.

\begin{figure}[ht]
\centerline{
\resizebox{9.0cm}{!}{\includegraphics{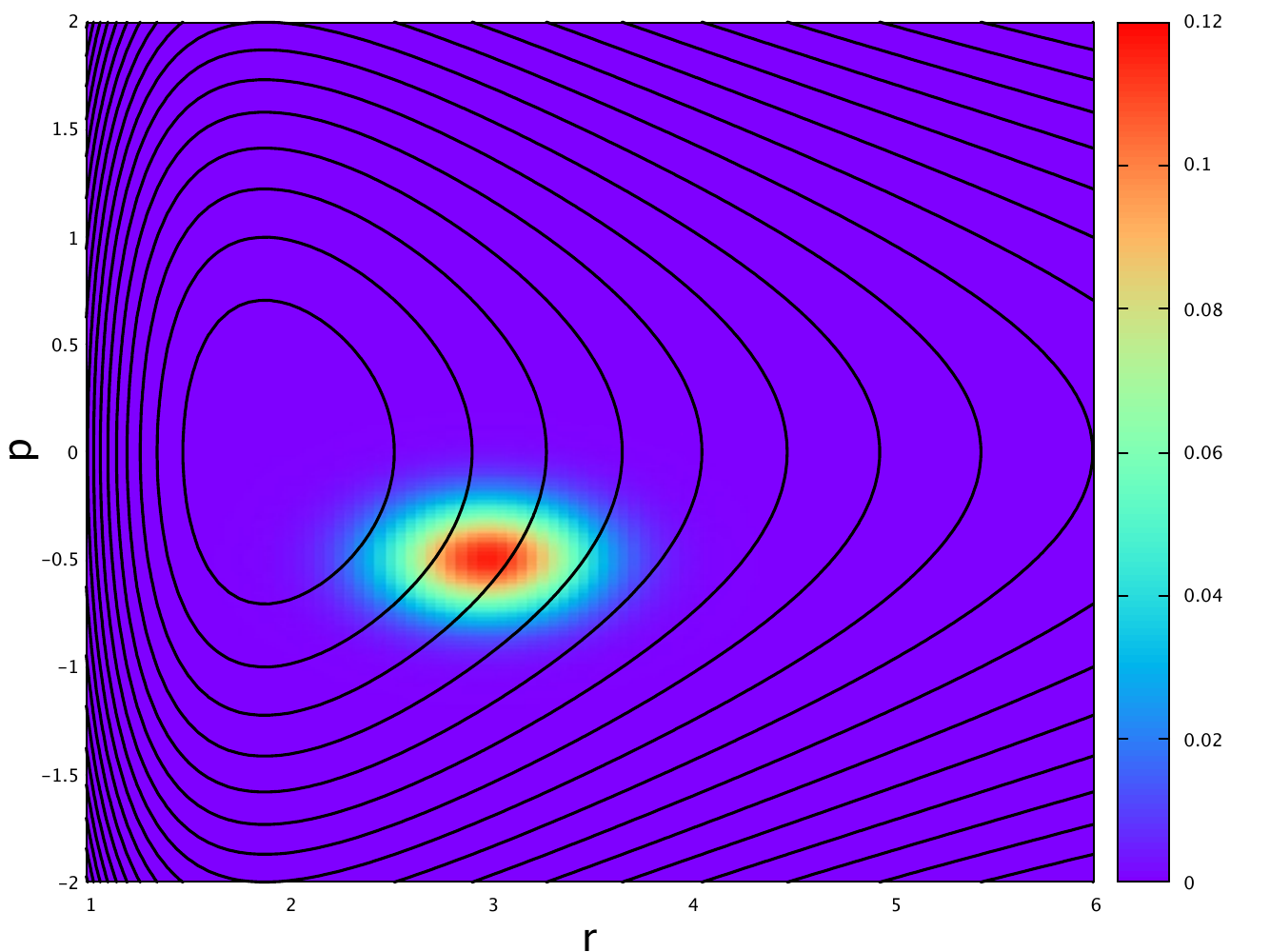}}
\resizebox{9.0cm}{!}{\includegraphics{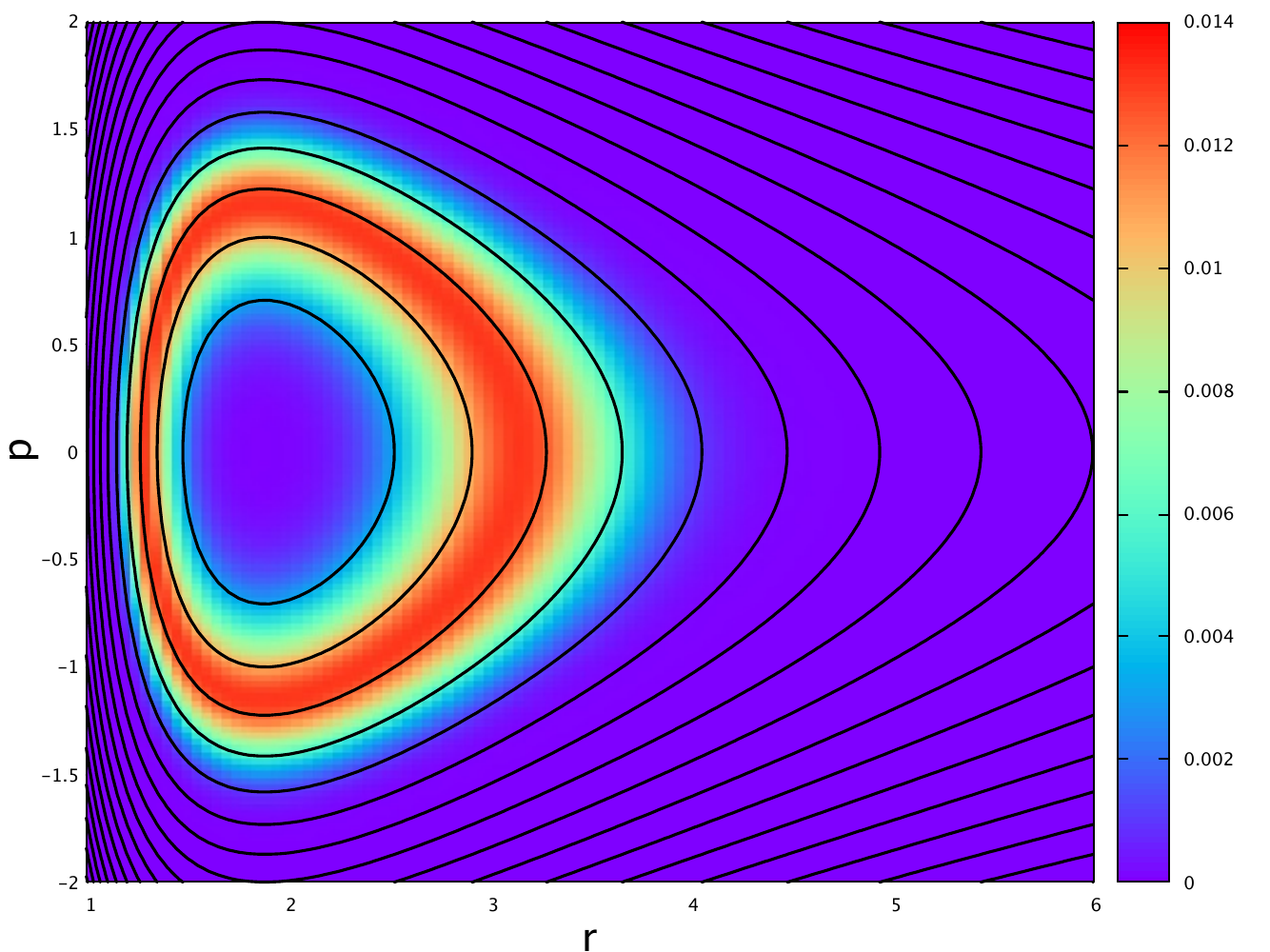}}}
\caption{\label{Fig:NFW} Contour plot of the initial (left panel) and averaged (right panel) distribution functions $f_0$ and $\langle f_0 \rangle_{C(E)}$ for the NFW potential. Also shown in these plots are the energy curves $C(E)$ which are indicated by the black contours. To obtain these plots we have also multiplied both $f_0$ and $\langle f_0 \rangle_{C(E)}$ with a factor of $4\pi$ to match the initial data used in the numerical simulation in~\cite{pDeJmAeMdN17}.
}
\end{figure}

In figure~\ref{Fig:NFW} we show the initial distribution function $f_0$ defined by Eq.~(\ref{Eq:InitialDistribution}) and its average $\langle f_0 \rangle_{(C(E)}$ for the NFW potential. The second plot should be compared with the lower-right panel of figure~17 in~\cite{pDeJmAeMdN17}. Although the results agree at the qualitative level, a more detailed comparison between $\langle f_0 \rangle_{(C(E)}$ and the numerical data for $f(t,\cdot)$ at the time of the final snapshot reveals some differences: the latter has a lower amplitude and a wider peak in its energy distribution, which presumably is due to numerical error and lack of resolution to resolve the mixed state (recall that the distribution function itself does not have a pointwise limit in phase space and develops large gradients; only the averaged quantities converge).

\subsection{Mixing inside each orbital plane}

In this section we consider a scenario in which the kinetic gas is confined to a fixed plane and is subject to one of the central potentials $\Phi(r)$ listed in Table~\ref{Tab:Potentials}, and we ask in which case mixing occurs. Hence, the problem is reduced to an effective two-dimensional problem for which Theorems~\ref{Thm:PSM3} and \ref{Thm:PSM4} apply, and mixing takes place if the condition~(\ref{Eq:NonDegeneracyCond2D}), involving the Hessian of the area function $A(E,L)$, is satisfied.

In the first case, the isochrone potential, it turns out the area function can be computed analytically (see~\cite{BinneyTremaine-Book} and Eq.~(\ref{Eq:AreaFunctionIsochrone}) in Appendix~\ref{App:AreaFunction}), with the result
\begin{equation}
A(E,L) = 2\pi\left[ \frac{1}{\sqrt{-2E}} - \frac{1}{2} \left( |L| + \sqrt{4 + L^2} \right) \right],
\qquad -\frac{2}{|L| + \sqrt{4 + L^2}} < E < 0,
\end{equation}
and based on the non-linear dependency on $E$ and $L$ it is simple to verify the validity of the determinant condition~(\ref{Eq:NonDegeneracyCond2D}). Therefore, mixing occurs for the isochrone potential.

\begin{figure}[ht]
\centerline{\resizebox{8.0cm}{!}{\includegraphics{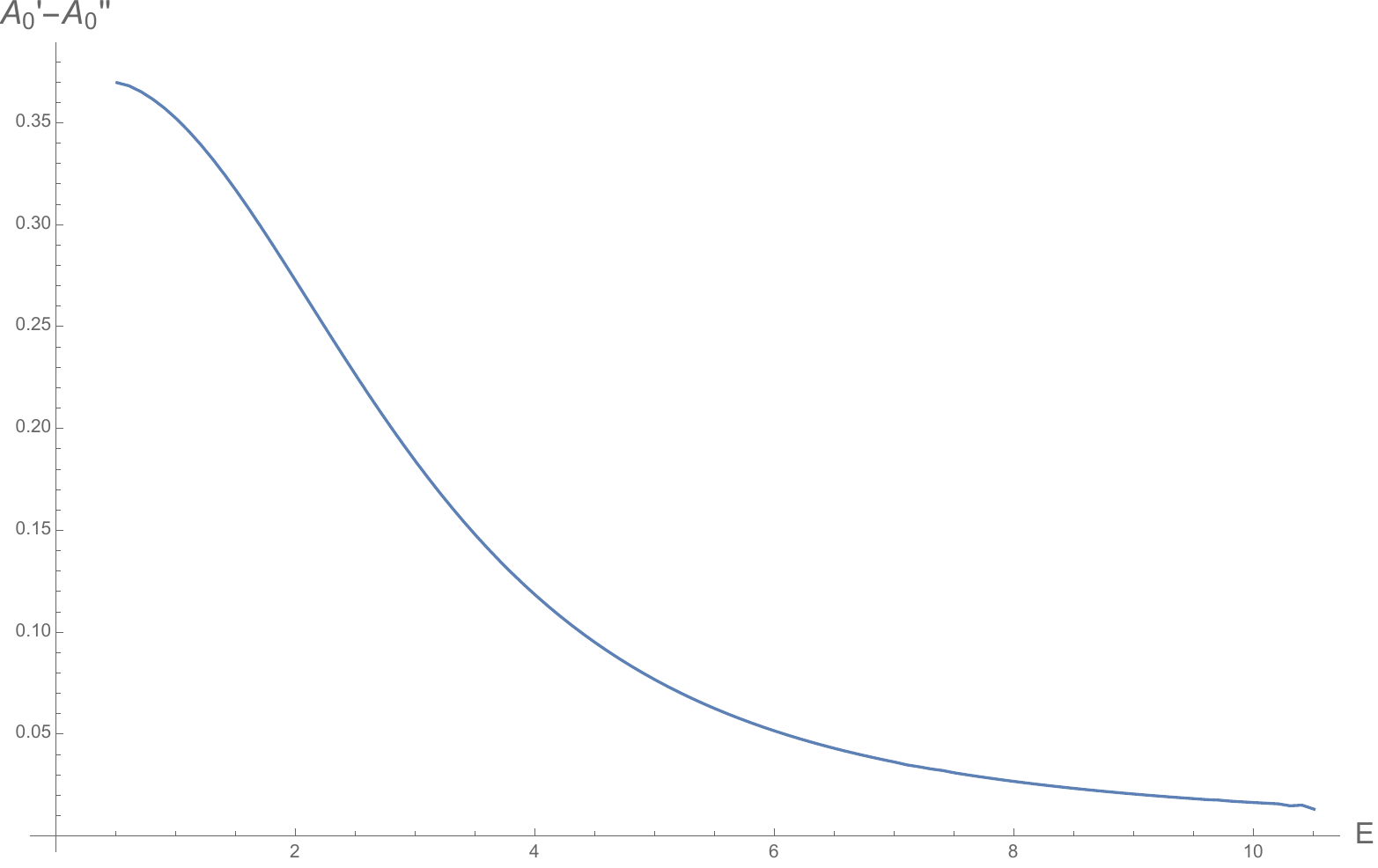}}}
\caption{\label{Fig:DA-DAA} A plot of the function $A_0' - A_0''$ that determines the sign of the determinant of $D^2 A(E,L)$ for the isothermal potential. As is visible from this plot, this quantity is positive, which, together with Eq.~(\ref{Eq:DetD2AIso}) and the fact that $A_0' > 0$ indicates that the determinant condition~(\ref{Eq:NonDegeneracyCond2D}) is satisfied.}
\end{figure}

Next, we analyze the case of the isothermal potential $\Phi(r) = 3\log(r)$. By setting $r = |L| x/\sqrt{3}$ we can rewrite the area function in the form
\begin{equation}
A(E,L) = 2|L|\int\limits_{x_1(E,L)}^{x_2(E,L)}
 \sqrt{2\left[ \frac{E}{3} - \log\left( \frac{|L|}{\sqrt{3}} \right) - \log(x) - \frac{1}{2x^2} \right]} dx 
  = |L| \left. A_0(\varepsilon) \right|_{\varepsilon = \frac{E}{3} - \log\left( \frac{|L|}{\sqrt{3}} \right) } \end{equation}
with
\begin{equation}
A_0(\varepsilon) := 2\int\limits_{x_1(\varepsilon)}^{x_2(\varepsilon)} 
\sqrt{2\left[ \varepsilon - \log(x) - \frac{1}{2x^2} \right]} dx,\qquad \varepsilon > \frac{1}{2},
\end{equation}
independent of $L$. Using this observation, the determinant of the Hessian of $A(E,L)$ yields
\begin{equation}
\det( D^2 A(E,L) ) = -\frac{1}{9} A_0'(\varepsilon)\left[  A_0'(\varepsilon) - A_0''(\varepsilon) 
 \right]_{\varepsilon = \frac{E}{3} - \log\left( \frac{|L|}{\sqrt{3}} \right) )}.
\label{Eq:DetD2AIso}
\end{equation}
We already know that $A_0'(\varepsilon) > 0$ and $A_0''(\varepsilon) > 0$ for all $\varepsilon > 1/2$ (see Lemma~\ref{Lem:Isothermal} in Appendix~\ref{App:PeriodFunction}); however, it remains to analyze the sign of $A_0' - A_0''$ in order to determine the one of $\det( D^2 A(E,L) )$. Although we were not able to come up with a rigorous simple proof for this fact, the plot in figure~\ref{Fig:DA-DAA} suggests that $A_0' - A_0'' > 0$, implying that the determinant condition is satisfied and mixing takes place also in the isothermal case.

\begin{figure}[ht]
\centerline{ \resizebox{7.0cm}{!}{\includegraphics{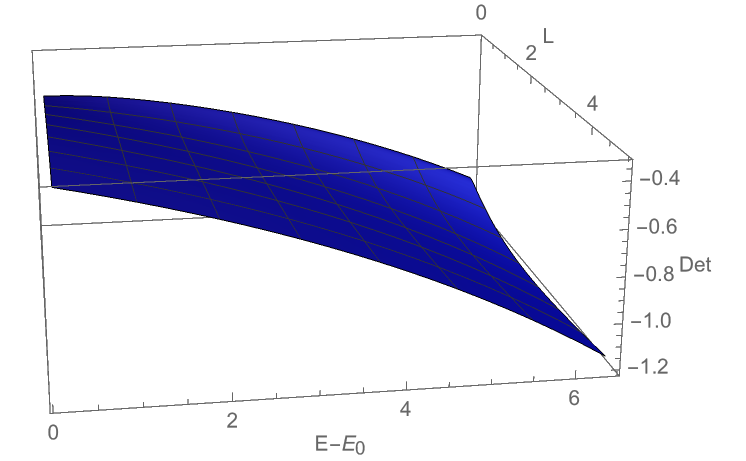}} \resizebox{7.0cm}{!}{\includegraphics{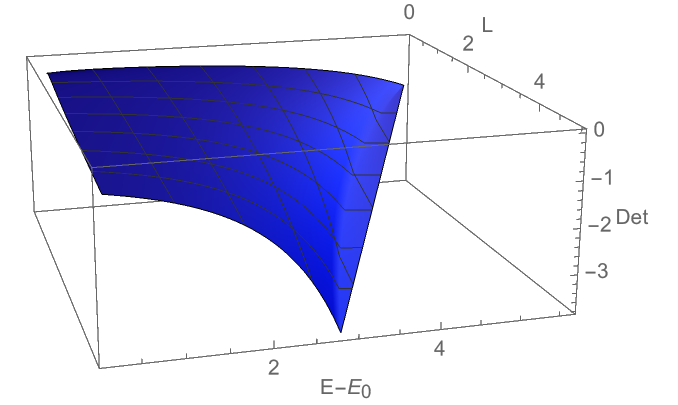}}  \resizebox{7.0cm}{!}{\includegraphics{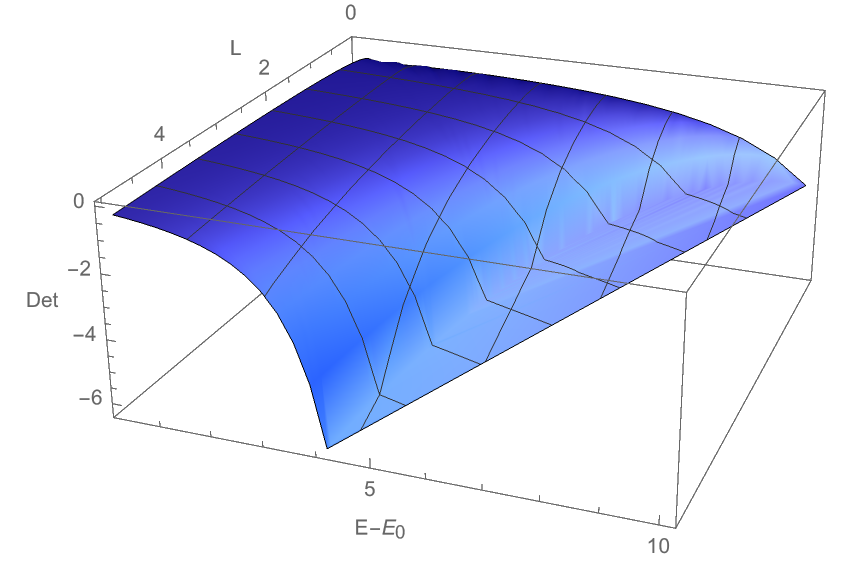}} }
\caption{\label{Fig:determin-models} Plots showing the determinant $\det( D^2 A(E,L) )$ as a function of $E-E_0$ and $L$, for the truncated isothermal, NFW and Burkert potentials (from left to right). These plots suggest that the determinant is negative, indicating that the determinant condition~(\ref{Eq:HessianH}) is satisfied.}
\end{figure}

Finally, in figure~\ref{Fig:determin-models} we provide plots for the determinant $\det( D^2 A(E,L) )$ in the remaining three case. These plots were obtained by numerically computing the Hessian $D^2 A(E,L)$ of the area function based on the formulae listed in Appendix~\ref{App:AreaFunction}. As these plots suggest, $\det( D^2 A(E,L) )$ is negative, such that mixing takes place (at least for the plotted range of parameters).

Summarizing, the results of this section show that the five potentials listed in Table~\ref{Tab:Potentials} exhibit the mixing phenomenon for a spherically symmetric configuration with fixed angular momentum $L$, as well as in the two-dimensional case of a gas confined to a specific plane. Furthermore, the comments in section~\ref{SubSec:3DModel} imply that in the full three-dimensional case, these potentials also yield mixing within each orbital plane, such that the final state is described by a distribution function depending only on the integrals of motion $(E,L_x,L_y,L_z)$.

%%%%%%%%%%%%%%%%%%%%%%%%%%%%%%%%%%%%%%%%%%%%%
\section{Application to the mixing properties of relativistic gas surrounding a static, spherically symmetric black hole}
\label{Sec:BlackHoles}
%%%%%%%%%%%%%%%%%%%%%%%%%%%%%%%%%%%%%%%%%%%%%

In this section, we generalize the results presented so far to the case of a collisionless relativistic gas trapped in the gravitational potential of a black hole spacetime. More specifically, we consider a general class of static, spherically symmetric black hole models consisting of a spacetime manifold ${\mathcal M}$ with metric of the form
\begin{equation}
ds^2 = -S(r)^2 N(r) dt^2 + \frac{dr^2}{N(r)} 
 + r^2\left( d\vartheta^2 + \sin^2\vartheta\, d\varphi^2 \right),
\label{Eq:Metric}
\end{equation}
with $S$ and $N$ given smooth functions of the areal radius $r$ which converge to one as $r\to \infty$ and are strictly positive for all $r > r_H$, while $N(r_H) = 0$ and $S(r_H) > 0$ at the event horizon radius $r = r_H$. In the Schwarzschild case one has $S = 1$ and $N(r) = 1 - r_H/r$; however for what follows the precise form of these coefficient will not be needed. We assume that the individual gas particles are confined to spatially bound, timelike geodesic trajectories in the exterior region $r > r_H$. For a detailed analysis of the unbounded case in a Schwarzschild spacetime with applications to accretion problems and a summary of the formulation of the relativistic Vlasov equation we refer the reader to our previous work in~\cite{pRoS17}. See also~\cite{pRoS18} for related results regarding bound orbits confined to the equatorial plane of a Kerr spacetime.

The Hamiltonian governing the motion of the gas particles is given by
\begin{equation}
{\mathcal H}(x,p) = \frac{1}{2} g^{\mu\nu}(x) p_\mu p_\nu
 = \frac{1}{2}\left[ \frac{-{p_{t}}^2}{S(r)^2N(r)} + N(r){p_r}^2+ \frac{1}{r^2}\left( p^2_{\vartheta}+\frac{p^2_{\varphi} }{ \sin^2 \vartheta} \right) \right],
\end{equation}
and besides ${\mathcal H}(x,p)$ itself, the system possesses the following integrals of motion:
\begin{equation}
{\mathcal E}(x,p) := -p_t,\quad
{\mathcal L}^2(x,p) :=p_{\vartheta} ^2 + \frac{p_{\varphi}^2}{\sin ^2 \vartheta},\quad
{\mathcal L}_z (x,p) := p_\varphi ,
\end{equation}
corresponding, respectively, to energy, total and azimuthal angular momentum. It is simple to verify that ${\mathcal H},{\mathcal E},{\mathcal L}^2,{\mathcal L}_z$ Poisson-commute with each other. The orbits are confined to the invariant subsets
\begin{equation}
\Gamma_{m,E,L, L_z} := \left\{ (x,p)\in T^*{\mathcal M} : \mathcal{H}(x,p) = \frac{m^2}{2}, {\mathcal E}(x,p) = E, {\mathcal L}^2(x,p) = L^2, {\mathcal L}_z (x,p) = L _z \right\},
\label{Eq:InvariantSet}
\end{equation}
of the cotangent space $T^*{\mathcal M}$ associated with ${\mathcal M}$. More explicitly, these sets are characterized by those coordinates $(t,r,\vartheta,\varphi,p_t,p_r,p_\vartheta,p_\varphi)$ of $T^*{\mathcal M}$ for which $p_t = -E$, $p_\varphi = L_z$ and $(r,p_r)$ and $(\vartheta,p_\vartheta)$ satisfy the equations
\begin{equation}
\left[ S(r) N(r) p_r \right]^2 + V_{L, m}(r) = E^2,\qquad
V_{L, m}(r) := S(r)^2 N(r)\left( m^2 + \frac{L ^2}{ r^2} \right),
\label{Eq:RadialMotion}
\end{equation}
and
\begin{equation}
 p_{\vartheta}^2 + \frac{L_z ^2}{\sin ^2 \vartheta} = L^2,
\label{Eq:PolarMotion}
\end{equation}
respectively. As mentioned before, in this work we focus on bound trajectories, and for this it is necessary to analyze the properties of the effective potential $V_{L, m}(r)$. To understand these properties, we need to further restrict our conditions on the functions $S$ and $N$ to ensure that $V_{L,m}$ behaves qualitatively similar to the Schwarzschild case in the exterior region $r > r_H$. We specify such conditions on the function $K(r) := S(r)^2 N(r)$ in Appendix~\ref{App:EffPotential}. They imply that the metric~(\ref{Eq:Metric}) describes a static, spherically symmetric and asymptotically flat black hole with a non-degenerate event horizon at $r = r_H$. Furthermore, these assumptions imply that the effective potential $V_{L,m}$ is monotonous for small enough values $L < L_{ms}$ of the total angular momentum, while for $L > L_{ms}$ the potential $V_{L,m}$ has a maximum at $r_{max}(L)\in (r_{ph},r_{ms})$ and a minimum at $r_{min}(L)\in (r_{ms},\infty)$, the radii $r_{ph}$ and $r_{ms}$ corresponding to those of the circular photon orbits and the innermost stable circular orbit, respectively. The range of parameters $(E,L,L_z)$ corresponding to bound orbits is given by the set
\begin{equation}
\Omega := \left\{ (E,L,L_z)\in \Real^3 : L > L_{ms}, E_{min}(L) < E < E_{max}(L), 0 < |L_z| < L \right\},
\end{equation}
with $E_{min}(L)$ and $E_{max}(L)$ defined in Eq.~(\ref{Eq:EminEmax}). ($E_{min}(L)$ represents the energy of the stable circular orbit $r = r_{min}(L)$ and $E_{max}(L)$ is either the energy of the unstable circular orbit $r = r_{max}(L)$ or $m$, whichever of the two is smaller.) For $m > 0$ and any $(E,L,L_z)\in \Omega$ one can easily verify that the invariant set $\Gamma_{m,E,L,L_z}$ defined in Eq.~(\ref{Eq:InvariantSet}) is a smooth submanifold of $T^*{\mathcal M}$ whose topology is $\Real \times T^3$. For the following, we focus on the phase space of bound trajectories $\Gamma_0$ consisting of the union of all these subspaces.

Next, we introduce action-angle like variables $(J_0,J_1,J_2,J_3,Q^0,Q^1,Q^2,Q^3)$ on $\Gamma_0$. Since $\Gamma_{m,E,L,L_z}$ is not compact, there is an ambiguity in the choice for $J_0$ corresponding to the non-compact dimension (see the discussion in section II.B in~\cite{tHeF08} and references therein). What seems to us the computationally simplest choice\footnote{For a different choice for $J_0$ which leads to an equivalent representation of the distribution function, see~\cite{tHeF08,pRoS18}.} for the action variables is obtained after the replacements $(m,E,L,L_z) \mapsto (\sqrt{-2{\mathcal H}(x,p)},{\mathcal E}(x,p),{\mathcal L}(x,p),{\mathcal L}_z(x,p))$ in the map
\begin{equation}
I_0(m,E,L,L_z) := m,\quad
I_1(m,E,L,L_z) := \frac{A_m(E,L)}{2\pi},\quad
I_2(m,E,L,L_z) : = L,\quad
I_3(m,E,L,L_z) := L_z,
\end{equation} 
which generalize the map~(\ref{Eq:IMap3d}) to the relativistic case. Here, the area function $A_m$ is defined by
\begin{equation}
A_m (E,L) := \oint p_r dr 
 = 2\int \limits_{r_1(E,L)}^{r_2(E,L)}\sqrt{E^2-V_{L,m}(r)} \frac{dr}{S(r)N(r)}, 
\end{equation}
with $r_1(E,L) < r_2(E,L)$ the turning points of the radial motion. The Jacobian of the map $I$ is given by
\begin{equation}
DI(m,E,L,L_z)  = \left(\begin{array}{cccc}
1 & 0 & 0 & 0 \\
\frac{1}{2\pi} \frac{\partial A_{m}(E,L)}{\partial m} & \frac{T(E,L)}{2\pi}  
 & -\frac{\Delta\Psi(E,L)}{2\pi} & 0\\
0 & 0 & 1 & 0 \\
0 & 0 & 0 & 1
\end{array} \right),
\label{Eq:JacobiMatrix4d}
\end{equation}
where the period function $T(E,L)$ of the radial motion and the phase shift function $\Delta\psi(E,L)$ are defined as
\begin{eqnarray}
T(E,L) &:=& \frac{\partial A_m}{\partial E}(E,L)
 = 2E \int \limits_{r_1(E,L)}^{r_2(E,L)}\frac{1}{\sqrt{E^2- V_{L,m} (r)}} \frac{dr}{S(r)N(r)},\\ 
\Delta\psi(E,L) &:=& -\frac{\partial A_m}{\partial L}(E,L) 
 = 2L \int \limits_{r_1(E,L)}^{r_2(E,L)}\frac{1}{\sqrt{E^2- V_{L,m} (r)}}\frac{S(r) dr}{r^2 },
\end{eqnarray}
and generalize the corresponding definitions~(\ref{Eq:TDef},\ref{Eq:DPsiDef}) to the relativistic case.\footnote{More precisely, $T(E,L)$ and $\Psi(E,L)$ represent the shifts in the coordinate $t$ and the angle $\psi$ in the orbital plane during which the radial coordinate moves from $r_1(E,L)$ to $r_2(E,L)$ and back, as can be verified by using the equations of motion $\dot{t} = E/(S^2 N)$, $\dot{r} = N p_r$ and $\dot{\psi} = L/r^2$.} In terms of the action variables, the Hamiltonian is simply ${\mathcal H}(x,p) = -J_0^2/2$, and thus the relativistic Vlasov equation reads
\begin{equation}
0 = \{ {\mathcal H} , f \} = -J_0\frac{\partial f}{\partial Q^0},
\end{equation}
which means that the one-particle distribution function should be a function depending only on $(Q^1,Q^2,Q^3,J_0,J_1,J_2,J_3)$. Here, the angle variables are obtained by differentiating the generating function
\begin{equation}
{\mathcal S}(x; I) := -E t + L_z\varphi + \int\limits_{r_1(E,L)}^r p_r dr 
 + \int\limits_{\pi/2}^\vartheta p_\vartheta d\vartheta 
\end{equation}
with respect to $I$, where the two integrals on the right-hand side should be understood as line integrals along the curves defined by Eqs.~(\ref{Eq:RadialMotion},\ref{Eq:PolarMotion}). Using the inverse transposed of the Jacobian in Eq.~(\ref{Eq:JacobiMatrix4d}) this yields the following expressions for the relevant angle variables $(Q^1,Q^2,Q^3)$ on which the distribution function depends:
\begin{eqnarray}
Q^1(x,p) &=& \left[ \frac{2\pi E}{T(E,L)} \int\limits_{r_1(E,L)}^r \frac{dr}{S^2 N^2 p_r} - \omega^1(E,L) t  \right]_{(E,L) = ({\mathcal E}(x,p),{\mathcal L}(x,p))},
\\
Q^2(x,p) &=& \left[ \frac{\Delta\psi(E,L) E}{T(E,L)} \int\limits_{r_1(E,L)}^r \frac{dr}{S^2 N^2 p_r}
 - L\int\limits_{r_1(E,L)}^r \frac{dr}{r^2 N p_r}
 + L\int\limits_{\pi/2}^\vartheta \frac{d\vartheta}{p_\vartheta} - \omega^2(E,L) t
 \right]_{(E,L,L_z) = ({\mathcal E}(x,p),{\mathcal L}(x,p),{\mathcal L}_z(x,p))},\\
Q^3(x,p) &=& \varphi - \left. L_z\int\limits_{\pi/2}^\vartheta \frac{d\vartheta}{\sin^2\vartheta p_\vartheta} \right|_{(L,L_z) = ({\mathcal L}(x,p),{\mathcal L}_z(x,p)},
\label{Eq:Q3}
\end{eqnarray}
with the frequencies $\omega^1(E,L) = 2\pi/T(E,L)$ and $\omega^2(E,L) = \Delta\psi(E,L)/T(E,L)$ having the same functional form as their Newtonian counterparts, see Eqs.~(\ref{Eq:omega_r},\ref{Eq:omega_phi}). It follows from the arguments presented in section~\ref{Sec:Main} that mixing in the angles $(Q^1,Q^2)$ takes place if the frequency function $\omega: \Omega_0\to \Real^2, (E,L)\mapsto (\omega^1(E,L),\omega^2(E,L))$ with $\Omega_0:=\{ (E,L)\in \Real^2 : L > L_{ms}, E_{min}(L) < E < E_{max}(L) \}$ satisfies
\begin{equation}
\det(D\omega(E,L)) \neq 0
\end{equation}
for almost all $(E,L)\in \Omega_0$. As in the Newtonian case, this condition can be reformulated as the following determinant condition for the Hessian of the area function $A_m(E,L)$:
\begin{equation}
\det( D^2 A_m (E,L) ) \neq 0,
\label{Eq:DetGeneralBH}
\end{equation}
for almost all $(E,L)\in \Omega_0$. In contrast to $(Q^1,Q^2)$, the angle variable $Q^3$ is constant in time and has the same interpretation as its Newtonian counterpart: it represents the angle between the $x$-axis and the intersection of the orbital plane with the $xy$-plane, as can be seen by setting $\vartheta = \pi/2$ in Eq.~(\ref{Eq:Q3}). Therefore, the integrals of motion $(Q^3,L,L_z)$ determine the three components of the angular momentum vector $(L_x,L_y,L_z)$ and the fulfillment of the determinant condition~(\ref{Eq:DetGeneralBH}) implies that the final state is described by a distribution function depending only on $(m,E,L_x,L_y,L_z)$.

The Hessian of the area function can be computed by generalizing the analysis presented in Appendix~\ref{App:AreaFunction} to the relativistic case. In many relevant situations it is even possible to express $A_m(E,L)$ and its derivatives in terms of special functions, which allows one to verify the determinant condition using a symbolic computing package. In particular, this is true for the Schwarzschild metric in which case $A_m (E,L)$ and its derivatives can be expressed in terms of complete elliptic integrals, see the results in~\cite{pRoS18} which also compute the area function $A_m(E,L)$ for equatorial bound orbits in a Kerr spacetime.

%%%%%%%%%%%%%%%%%%%%%%%%%%%%%%%%%%%%%%%%%%%%
\section{Conclusions and future work}
\label{Sec:Conclusions}
%%%%%%%%%%%%%%%%%%%%%%%%%%%%%%%%%%%%%%%%%%%%

In this work, we have provided a thorough discussion of phase space mixing for a collisionless kinetic gas which is trapped in an external, $d$-dimensional, rotationally symmetric potential well. As discussed in the introduction, phase space mixing implies that the macroscopic observables converge in time to those belonging to an ``equilibrium" distribution function which depends only on integrals of motion and which can be computed from a suitable average of the initial distribution function. The main results of this article, whose precise statements are described in the collection of theorems in section~\ref{Sec:Main} which are proven in Appendix~\ref{App:Proofs} can be summarized as follows: in the one-dimensional case ($d=1$) mixing occurs if the period function $T(E)$ (describing the period of the orbit with energy $E$) has a first derivative which is almost everywhere different from zero (thus excluding the case of the harmonic oscillator for which $T$ is constant). In $d = 2,3$ dimensions mixing occurs if the area function $A(E,L)$, describing the area enclosed by the orbit with energy $E$ and (total) angular momentum $L$ in phase space $(r,p_r)$ describing the radial motion, has a Hessian which is invertible for almost all admissible values of $(E,L)$ parametrizing bound orbits. While this condition excludes the harmonic and Kepler potentials (for which all bound orbits are closed), it is satisfied for many interesting examples as we have shown in this article. For $d=2$ the resulting equilibrium distribution function depends on $(E,L)$ only, while for $d=3$ it is a function of $E$ and the three components of the angular momentum vector $\vec{L}$. Thus, the final state is compatible with the strong Jeans theorem which states that the distribution function describing a steady state in which almost all orbits are regular can be written as a function of isolating integrals of motion~\cite{dL62a,MoBoschWhite-Book}. However, our results go beyond this theorem by proving that (at least in the systems considered in this article) any kinetic gas configuration will converge in time to an equilibrium configuration described by such a steady state.

To provide examples and applications, we have analyzed a family of potentials which are relevant in stellar dynamics and the modeling of dark matter halos, and we have shown that each of these potentials satisfy the required conditions on the area function $A(E,L)$ for mixing to take place in $d = 2,3$ dimensions. Further, we have shown that phase space mixing occurs for a spherically symmetric gas configuration in which all the gas particles have the same value of the total angular momentum number $L > 0$ (and hence behaves like an effective one-dimensional system). This provides a qualitative understanding for the virialization phenomenon observed in~\cite{pDeJmAeMdN17} through phase space mixing, and offers a simple method for computing the distribution function associated with the final state from an angle average of the initial distribution function.

We have also generalized our results to the general-relativistic setting of a collisioness gas whose individual gas particles follow bound timelike geodesic trajectories in a static, spherically symmetric black hole spacetime. It has been shown that phase space mixing (with respect to static observers) takes place provided a condition on the area function is satisfied which is analogous to the Newtonian case. The results from our previous work~\cite{pRoS18}, besides offering a coordinate-independent description of the mixing phenomenon, suggest that the condition on $A(E,L)$ is satisfied for a Schwarzschild black hole. In fact, the results presented in~\cite{pRoS18} also provide a generalization of this condition for equatorial orbits in a (rotating) Kerr black hole, and they show that a thin equatorial disk made out of a collisionless gas is subject to phase space mixing, the final state being stationary and axisymmetric. It should be mentioned that the proof given in~\cite{pRoS18} required the determinant of the Hessian of $A(E,L)$ to be different from zero for all $(E,L)$ intersecting the support of the test function. This requirement led to a non-trivial restriction on the support of the test function, since it has also been observed in~\cite{pRoS18} that the determinant is zero for a certain zero-measure set $Z$ in the space of admissible $(E,L)\in \Omega_0$ leading to bound orbits. However, the  results of the present article show that mixing still takes place if the determinant is everywhere non-vanishing except on such a set of measure zero in $\Omega_0$, and hence they are expected to strengthen our results in~\cite{pRoS18} by removing the requirement on the support of the test function.

In future work, we plan to analyze more general gas configurations on a Kerr spacetime, which are not necessarily confined to its equatorial plane. Since the geodesic flow on a Kerr background is described by an integrable Hamiltonian system~\cite{bC68,mWrP70}, this can in principle be analyzed with the same techniques as the ones used in the present article. Further interesting and relevant questions we have left open and hope to come back to in future work concern the analysis of the time scale on which the relaxation process takes place and the generalization to the self-gravitating case, where one needs to consider either the coupled Vlasov-Poisson or the coupled Vlasov-Einstein system of equations.

We conclude this article by pointing out that in recent years there has been a considerable interest in mathematical studies of the Vlasov-Einstein system, see the review article~\cite{hA11} and Refs.~\cite{dFjJjS15,dFjJjS17,mT16,hLmT20,lBdFjJjSmT20} for recent work regarding the nonlinear stability of the Minkowski spacetime in this system. In relation to our application of the mixing phenomenon to black hole spacetimes, it is interesting to note that the works in~\cite{lApBjS18,lB20} prove time decay for the solutions of the \emph{massless} Vlasov equation on a Schwarzschild or slowly rotating Kerr background. The decay in the massless case is related to the fact that in the Schwarzschild exterior, null geodesic bound orbits only occur for the special radius $r = 3r_H/2$ and are unstable, leading to dispersion. In contrast to this, stable timelike geodesic bound orbits exist for a whole range of radii, such that a kinetic gas configuration of massive particles does not disperse completely, but converges to a non-trivial steady state. Therefore, it is expected that phase space mixing will play a prominent role when establishing rigorous results regarding the asymptotic behavior of dynamical solutions to the massive Vlasov-Einstein system whose initial data do not necessarily lie close to Minkowski data.

%%%%%%%%%%%%%%%%%%%%%%%%%%%
%%%   ACKNOWLEDGMENTS
%%%%%%%%%%%%%%%%%%%%%%%%%%%

\acknowledgments

It is a pleasure to thank Carlos Gabarrete, Ulises Nucamendi, Dar\'io N\'u\~nez, and Thomas Zannias for fruitful and stimulating discussions. We also thank Piotr Chru\'sciel for pointing out to us Refs.~\cite{sCdW86,cC87}, Jim Ellison for sending to us Ref.~\cite{cM19}, and Petr Zhevandrov for drawing our attention to Ref.~\cite{CornfeldFominSinai-Book}. Finally, we thank Paola Dom\'inguez and Erik Jim\'enez for providing to us the numerical data obtained in~\cite{pDeJmAeMdN17} and an anonymous referee for pointing out to us the connection of our work with those of Ref.~\cite{lApBjS18,lB20}. This research was supported in part by CONACyT Grants No.~577742, by CONACyT Network Project No.~280908 ``Agujeros Negros y Ondas Gravitatorias" and by a CIC Grant to Universidad Michoacana.

\appendix

%%%%%%%%%%%%%%%%%%%%%%%%%%%%%%%%%%%%%%%%%%%%%
\section{Proofs of the main theorems in section~\ref{Sec:Main}}
\label{App:Proofs}
%%%%%%%%%%%%%%%%%%%%%%%%%%%%%%%%%%%%%%%%%%%%%

This appendix is devoted to the proofs of the main theorems stated in section~\ref{Sec:Main}. These proofs are based on the explicit solution representations in terms of action-angle variables discussed in the last part of that section and use the Fourier transform in the angle variables. One then shows that, due to the Riemann-Lebesgue lemma, all Fourier coefficients  decay to zero as $t\to \infty$, except the zero mode which is related to the angle average. The idea of this proof can already be found in the paper by Lynden-Bell~\cite{dL62} on stellar dynamics. Our proof is mostly based on a very recent paper by Mitchell~\cite{cM19} which provides all the necessary tools for the proofs of Theorems~\ref{Thm:PSM1} and \ref{Thm:PSM3}. For completeness, we summarize these tools in this appendix and show that they can be used and generalized to prove the main theorems~\ref{Thm:PSM1},\ref{Thm:PSM2},\ref{Thm:PSM3},\ref{Thm:PSM4}. To our knowledge, the proofs of Theorems~\ref{Thm:PSM2} and~\ref{Thm:PSM4} for the distributional test functions are new.

It is also possible to provide an alternative proof for Theorem~\ref{Thm:PSM1} which is based on a two-dimensional Fourier transformation and parallels the standard proof of phase space mixing for a free gas on a torus, see for example Ref.~\cite{cMcV11}, section~3. We will only make brief comments at the end of this appendix regarding this alternative proof.

\subsection{Formal calculation and idea of the proof}

Recall from section~\ref{SubSec:AA} that on the subset $\Gamma_0$ of phase space, the one-particle distribution function $f$ can be represented in the following explicit form:
\begin{equation}
f(t,x,p) = F\left( Q(x,p) - \omega(J(x,p))t, J(x,p) \right),\qquad (x,p)\in \Gamma_0,
\label{Eq:ExplicitSolution}
\end{equation}
with $(Q,J)\in T^d\times \Omega_J$ the action-angle coordinates (which are smooth functions on $\Gamma_0$ as discussed in section~\ref{SubSec:AA}), and where $F: T^d\times \Omega_J\to \Real$ is the action-angle representation of the initial datum $f_0\in L^1(\Gamma)$, i.e.
\begin{equation}
f_0(x,p) = F(Q(x,p),J(x,p)),\qquad (x,p)\in \Gamma_0.
\end{equation}
Here and in the following, we treat the $(d=1)$ and $(d=2)$-dimensional cases on the same footing, where for $d=1$, $\Omega_J = (0,\infty)$ while for $d=2$ the open subset $\Omega_J$ of $\Real^2$ is defined below Eq.~(\ref{Eq:DPsiDef}).

Let $\Psi: T^d\times\Omega_J\to \Real$ be the action-angle representation of the test function $\varphi$, such that
\begin{equation}
\varphi(x,p) = \Psi(Q(x,p),J(x,p)),\qquad (x,p)\in \Gamma_0.
\end{equation}
Because the transformation $(x,p)\mapsto (Q,J)$ is symplectic and hence volume-preserving, we have
\begin{equation}
N_\varphi(t) = \int\limits_{\Gamma_0} f(t,x,p)\varphi(x,p) d^dx d^d p
 = \int\limits_{\Omega_J} \int\limits_{T^d} F_t(Q,J) \Psi(Q,J) d^d Q d^d J,
\label{Eq:NFPsi}
\end{equation}
with
\begin{equation}
F_t(Q,J) := F(Q - \omega(J) t,J).
\label{Eq:FrequencyRepresentation}
\end{equation}

Next, we represent the functions $F$ and $\Psi$ in terms of their Fourier series with respect to the angle variables $Q$. Let
\begin{equation}
\hat{F}_k(J) := \frac{1}{(2\pi)^{d/2}} \int\limits_{T^d} F(Q,J) e^{-i k\cdot Q} d^d Q,
\qquad k\in \Integer^d, \quad J\in \Omega_J,
\end{equation}
be the Fourier coefficients of $F$, and likewise for $\hat{\Psi}_k$. Noting that the Fourier coefficients of $F_t$ are given by $\hat{F}_k(J) e^{-i k\cdot\omega(J) t}$ and using Parseval's identity one obtains
\begin{equation}
N_\varphi(t) = \int\limits_{\Omega_J} \sum\limits_{k\in \Integer^d} 
\hat{F}_k(J)\hat{\Psi}_k(J)^* e^{-i k\cdot\omega(J) t}  d^d J,
\label{Eq:NFhatPsihat}
\end{equation}
with $\hat{\Psi}_k(J)^*$ denoting the complex conjugate of $\hat{\Psi}_k(J)$. Let
\begin{equation}
h_k(J) := \hat{F}_k(J)\hat{\Psi}_k(J)^*,\qquad k\in \Integer^d,\quad J\in \Omega_J.
\end{equation}
We shall establish the following key properties:
\begin{enumerate}
\item[(a)] $h_k\in L^1(\Omega_J)$ for each $k\in \Integer^d$ and
\begin{equation}
\sum\limits_{k\in \Integer^d}  \| h_k \|_{L^1(\Omega_J)} < \infty,\qquad
\| h_k \|_{L^1(\Omega_J)} := \int\limits_{\Omega_J} |h_k(J)| d^d J.
\end{equation}
\item[(b)] For all $k\inÊ\Integer^d\setminus \{ 0 \}$,
\begin{equation}
\lim\limits_{t\to \infty} \int\limits_{\Omega_J} h_k(J) e^{-i k\cdot\omega(J) t} d^dJ = 0.
\end{equation}
\item[(c)] Under the hypothesis of the theorems, the equality~(\ref{Eq:NFhatPsihat}) holds for all $t\geq 0$ (in a strict, not just formal, sense).
\end{enumerate}

Once these properties have been established, the mixing property follows immediately: properties (a) and (c) allow one to write
\begin{equation}
\lim\limits_{t\to \infty} N_\varphi(t) = \sum\limits_{k\in \Integer^d}  \lim\limits_{t\to \infty}
  \int\limits_{\Omega_J} h_k(J) e^{-i k\cdot\omega(J) t} d^dJ.
\end{equation}
According to property (b), only the $k = 0$ term survives in the sum over $k$, and thus
\begin{eqnarray}
\lim\limits_{t\to \infty} N_\varphi(t) 
 &=& \int\limits_{\Omega_J} \hat{F}_0(J)\hat{\Psi}_0(J)^* d^d J
\nonumber\\
 &=& \frac{1}{(2\pi)^d} \int\limits_{\Omega_J} \left( \int\limits_{T^d} F(Q,J) d^dQ \right)
 \left( \int\limits_{T^d} \Psi(Q,J) d^d Q\right) d^d J
\nonumber\\
 &=& \int\limits_{\Omega_J}\int\limits_{T^d} \langle F \rangle(Q,J) \Psi(Q,J) d^d Q d^d J,
\end{eqnarray}
where in the last step we have used the definition of the angle-average of $F$, see Eqs.~(\ref{Eq:CEAverage}) and (\ref{Eq:AngleAverage}).

Therefore, all that remains to be done is to establish the key properties (a), (b) and (c). In the next subsections, we will show that the hypotheses of Theorems~\ref{Thm:PSM1}--\ref{Thm:PSM4} imply the satisfaction of these properties for a dense subset of initial conditions $F$, which allows to prove the theorems.

\subsection{The generalized Riemann-Lebesgue lemma and property (b)}

We start with property (b), which is a generalization of the Riemann-Lebesgue lemma, cf. Lemma~0 in~\cite{cM19}.

\begin{lemma}
\label{Lem:RiemannLebesgue}
Let $\Omega\in \Real^d$ be an open subset of $\Real^d$, let $h\in L^1(\Omega,\Complex)$ and $w\in C^2(\Omega,\Real)$ be such that for almost all $x\in \Omega$,
\begin{equation}
\nabla w(x) \neq 0.
\end{equation}
Then,
\begin{equation}
\lim\limits_{|t|\to \infty} \int\limits_\Omega h(x) e^{-i w(x) t} d^d x = 0.
\end{equation}
\end{lemma}

\proof Since $\nabla w:\Omega \to \Real^d$ is continuous, the set
\begin{equation}
Z := \{ x\in \Omega : \nabla w(x) = 0 \}
\end{equation}
is closed, and it is also a zero-measure set by assumption. Therefore,
\begin{equation}
 \int\limits_\Omega h(x) e^{-i w(x) t} d^d x 
 = \int\limits_{\Omega\setminus Z}  h(x) e^{-i w(x) t} d^d x,
\end{equation}
and it is sufficient to establish the theorem replacing $\Omega$ with the open set $\Omega_0 := \Omega\setminus Z$, on which $w$ has no critical points.

Next, we assume that $h\in C^\infty_0(\Omega_0)$ is smooth and compactly supported in $\Omega_0$. Introduce the vector field $X: \Omega_0\to \Real^d$ defined by
\begin{equation}
X(x) := h(x)\frac{\nabla w(x)}{|\nabla w(x)|^2},\qquad x\in \Omega_0.
\label{Eq:XDef}
\end{equation}
For all $t\neq 0$ it follows that
\begin{equation}
\int\limits_{\Omega_0} h(x) e^{-i w(x) t} d^d x 
 = -\frac{1}{it}\int\limits_{\Omega_0} X(x)\cdot \nabla\left( e^{-i w(x) t} \right) d^d x
 = \frac{1}{it}\int\limits_{\Omega_0} (\nabla\cdot X)(x) e^{-i w(x) t} d^d x,
\end{equation}
where in the last step we have used integration by parts and the fact that $X$ is compactly supported in $\Omega_0$. According to the assumptions, $\nabla\cdot X$ is continuous, and hence
\begin{equation}
\left| \int\limits_{\Omega_0} h(x) e^{-i w(x) t} d^d x  \right| 
 \leq \frac{1}{|t|} \int\limits_{\Omega_0} | \nabla\cdot X(x) | d^d x\to 0,
\end{equation}
as $|t|\to \infty$. This proves the lemma for $h\in C^\infty_0(\Omega_0)$.

Finally, if $h\in L^1(\Omega_0)$, we use the fact that $C_0^\infty(\Omega_0)$ is dense in $L^1(\Omega_0)$ to approximate $h$ by a sequence $h_n$ in $C^\infty_0(\Omega_0)$ such that $h_n\to h$ in $L^1(\Omega_0)$. Then,
\begin{eqnarray}
\left| \int\limits_{\Omega_0} h(x) e^{-i w(x) t} d^d x  \right| 
 &\leq& \int\limits_{\Omega_0} | h(x) - h_n(x) | d^d x 
 + \left| \int\limits_{\Omega_0} h_n(x) e^{-i w(x) t} d^d x \right| 
\nonumber\\
 &\leq& \| h - h_n \|_{L^1(\Omega)} 
  + \left| \int\limits_{\Omega_0} h_n(x) e^{-i w(x) t} d^d x \right|.
\end{eqnarray}
Given $\varepsilon > 0$ we first choose $n$ large enough such that $ \| h - h_n \|_{L^1(\Omega)} < \varepsilon/2$. Then, by the result of the first part of the proof one can choose $t_0 > 0$ large enough such that the second term on the right-hand side is smaller than $\varepsilon/2$ for all $|t| > t_0$. It then follows that
\begin{equation}
\left| \int\limits_{\Omega_0} h(x) e^{-i w(x) t} d^d x  \right| < \varepsilon,\qquad
|t| > t_0,
\end{equation}
which establishes the statement also for $h\in L^1(\Omega)$.
\qed

{\bf Remark}:  It follows from the proof that if the vector field $X: \Omega_0\to \Real^d$ defined in Eq.~(\ref{Eq:XDef}) has only removable singularities and can be extended to a vector field $\overline{X}:\Omega\to \Real^d$ that lies in the class $H^{1,1}_0(\Omega,\Real^d)$, then one obtains $1/t$ decay:
\begin{equation}
\left| \int\limits_{\Omega_0} h(x) e^{-i w(x) t} d^d x  \right|
  \leq \frac{1}{|t|} \| \nabla\cdot\overline{X} \|_{L^1(\Omega)}.
\end{equation}

Assuming that $h_k\in L^1(\Omega_J)$, the previous lemma implies property (b). Indeed, for the setting of Theorems~\ref{Thm:PSM1}--\ref{Thm:PSM4}, $\Omega_J$ is open and the frequency function $\omega: \Omega_J\to \Real^2$ is smooth and satisfies
\begin{equation}
\det( D\omega(J) ) \neq 0
\end{equation}
for almost all $J\in \Omega_J$. For $k\in \Integer^d\setminus \{ 0 \}$ it then follows that the function
\begin{equation}
w(J) := k\cdot\omega(J),\qquad J\in \Omega_J
\end{equation}
is smooth and that its gradient $\nabla w(J) = k\cdot (D\omega(J))$ must be different from zero for almost all $J\in \Omega_J$, since otherwise $D\omega(J)$ would have a non-trivial kernel. Therefore, Lemma~\ref{Lem:RiemannLebesgue} implies property (b) provided $h_k\in L^1(\Omega_J)$ for all $k\in \Integer^d$.

\subsection{Proofs of Theorems~\ref{Thm:PSM1} and \ref{Thm:PSM3}}

After these remarks and preliminary results, it is not difficult to complete the proofs of Theorem~\ref{Thm:PSM1} and its two-dimensional generalization Theorem~\ref{Thm:PSM3}. We assume first that $F,\Psi\in C_b(T^d\times\Omega_J)$ are bounded, continuous functions and that $F$ has compact support, such that $F(Q,J) = 0$ for all $J\in \Omega_J\setminus K$ outside a compact subset $K$ of $\Omega_J$. It follows by Lebesgue's dominated convergence theorem that for each $k\in \Integer^d$, the functions $\hat{F}_k, \hat{\Psi}_k: \Omega_J\to \Complex$ are continuous and bounded. In particular, it follows that $h_k := \hat{F}_k \hat{\Psi}_k^*\in L^1(\Omega_J)$ since $\hat{F}_k$ vanishes outside $K$. Furthermore, by Parseval's identity
\begin{equation}
\sum\limits_{k\in \Integer^d} | \hat{F}_k(J)|^2 = \int\limits_{T^d} |F(Q,J)|^2 d^d Q,\qquad
\sum\limits_{k\in \Integer^d} | \hat{\Psi}_k(J)|^2 = \int\limits_{T^d} |\Psi(Q,J)|^2 d^d Q,
\end{equation}
for all $J\in \Omega_J$. Using the Cauchy-Schwarz inequality it follows that
\begin{equation}
\sum\limits_{k\in \Integer^d} |h_k(J)| \leq
\left( \int\limits_{T^d} |F(Q,J)|^2 d^d Q \right)^{1/2}
\left( \int\limits_{T^d} |\Psi(Q,J)|^2 d^d Q \right)^{1/2}
\leq (2\pi)^d M_1 M_2 \chi_K(J),
\end{equation}
where $M_1 := \max\{ |F(Q,J)| : (Q,J)\in T^d\times\Omega_J \}$, $M_2 := \max\{ |\Psi(Q,J)| : (Q,J)\in T^d\times\Omega_J \}$ and $\chi_K(J)$ is the indicator function of the set $K$ (i.e. $\chi_K(J) = 1$ for $J\in K$ and $\chi_K(J) = 0$ otherwise). Since $K\subset\Omega_J$ is compact, property (a) follows. Using Parseval's identity again reveals that
\begin{equation}
\int\limits_{T^d} F_t(Q,J)\Psi(Q,J) d^d Q 
 = \sum\limits_{k\in \Integer^d} \hat{F}_k(J)\hat{\Psi}_k(J)^* e^{-i k \omega(J) t}
\end{equation}
for all $J\in \Omega_J$, and because of property (a) we can integrate both sides over $J$, obtaining Eq.~(\ref{Eq:NFhatPsihat}), such that also property (c) holds. This proves the theorem for $F\in C_0(T^d\times\Omega_J)$ continuous with compact support.

Finally, let $F\in L^1(T^d\times\Omega_J)$.\footnote{Note that because of the identity
$$
\int\limits_{\Omega_J}\int\limits_{T^d} F(Q,J) d^d Q d^d J = \int\limits_{\Gamma_0} f_0(x,p) d^d x d^d p
$$
it follows that $F\in L^1(T^d\times\Omega_J)$ if and only if $f_0\in L^1(\Gamma_0)$. Likewise, $F\in C_b(T^d\times\Omega_J)$ is continuous and bounded if and only if $f_0\in C_b(\Gamma_0)$ is.} Then, we may approximate $F_n\to F$ in $L^1(T^d\times\Omega_J)$ by functions $F_n\in C_0(T^d\times\Omega_J)$ which are continuous and have compact support. Using this, we obtain
\begin{eqnarray}
&& \left| \int\limits_{\Omega_J}\int\limits_{T^d}
 \left[ F_t(Q,J) - \langle F \rangle(Q,J) \right]\Psi(Q,J) d^d Q d^d J \right|
\nonumber\\
 &&\qquad = \left| \int\limits_{\Omega_J}\int\limits_{T^d}
 \left[ F(Q,J) - \langle F \rangle(Q,J) \right]\Psi_{-t}(Q,J) d^d Q d^d J \right|
\nonumber\\
 &&\qquad \leq \int\limits_{\Omega_J}\int\limits_{T^d}
 | F(Q,J) - F_n(Q,J) | |\Psi_{-t}(Q,J)| d^d Q d^d J
  + \int\limits_{\Omega_J}\int\limits_{T^d}
 | \langle F_n\rangle(Q,J) - \langle F \rangle(Q,J) | |\Psi_{-t}(Q,J)| d^d Q d^d J
 \nonumber\\
  &&\qquad + \left| \int\limits_{\Omega_J}\int\limits_{T^d}
 \left[ F_n(Q,J) - \langle F_n \rangle(Q,J) \right]\Psi_{-t}(Q,J) d^d Q d^d J \right|
\nonumber\\
 && \qquad \leq 2M_2 \| F_n - F \|_{L^1(T^d\times\Omega_J)} 
  + \left| \int\limits_{\Omega_J}\int\limits_{T^d}
 \left[ F_n(Q,J) - \langle F_n \rangle(Q,J) \right]\Psi_{-t}(Q,J) d^d Q d^d J \right|,
\end{eqnarray}
where we have used the variable substitution $Q\mapsto Q + \omega(J) t$ and the definition $\Psi_{-t}(Q,J) := \Psi(Q + \omega(J)t,J)$ in the first step. Given $\varepsilon > 0$, we may choose $n$ large enough, such that $\| F_n - F \|_{L^1(T^d\times\Omega_J)} \leq \varepsilon/(4M_2)$ and then $t_0 > 0$ large enough such that the last term is smaller than $\varepsilon/2$ for all $t > t_0$. This concludes the proof of Theorems~\ref{Thm:PSM1} and \ref{Thm:PSM3}.

\subsection{Proof of Theorem~\ref{Thm:PSM2}}

Next, we prove Theorem~\ref{Thm:PSM2} which corresponds to a distributional test function of the form
\begin{equation}
\varphi(x,p) = \delta(x - x_0) g(p),
\end{equation}
with $g: \Real\to \Real$ continuous and $x_0\in I$ fixed. Although $\varphi$ and the corresponding test function $\Psi(Q,J)$ in terms of action-angle variables are distributional, their Fourier coefficients are regular:
\begin{equation}
\hat{\Psi}_k(J) = \frac{1}{\sqrt{2\pi}}\frac{\omega(J)}{p} 
\left[ g(p) e^{-i k Q(x_0,p)} + g(-p) e^{i k Q(x_0,p)} \right]_{p = \sqrt{2(H(J) - V(x_0))}}
\Theta\left[ H(J) - V(x_0) \right],
\label{Eq:PsiHatk}
\end{equation}
where $H(J)$ is the energy as a function of the action variable $J$, $\omega(J) = dH(J)/dJ$ the frequency function, $\Theta(u)$ denotes the Heaviside function which is one if $u > 0$ and zero for $u \leq 0$, and $Q(x,p)$ is the angle variable defined in Eq.~(\ref{Eq:QDef}) which satisfies $dQ/dx = \omega(J)/p$ along the energy curve $C(E)$. The two terms on the right-hand side of Eq.~(\ref{Eq:PsiHatk}) arise due to the fact that the line $x = x_0$ in phase space intersects the energy curve $C(E)$ at two points $(x_0,\pm p)\in C(E)$ when $E > V(x_0)$, corresponding to two angles $Q(x_0,p)$ and $Q(x_0,-p)$ such that $Q(x_0,p) + Q(x_0,-p) = 2\pi$.

Eq.~(\ref{Eq:PsiHatk}) immediately yields the estimate
\begin{equation}
|\hat{\Psi}_k(J)| \leq \frac{1}{\sqrt{2\pi}}\frac{\omega(J)}{p}
\left[ |g(p) + |g(-p)| \right]_{p = \sqrt{2(H(J) - V(x_0))}}\Theta\left[ H(J) - V(x_0) \right].
\label{Eq:PsiHatkBound}
\end{equation}
The next lemma provides sufficient conditions for the satisfaction of the key properties (a) and (c).

\begin{lemma}
\label{Lem:Fck}
Suppose $F: S^1\times (0,\infty) \to \Real$ is continuously differentiable, and suppose there exists a non-negative continuous function $P: (0,\infty)\to \Real$ such that
\begin{equation}
\int\limits_{-\infty}^\infty P( J(x_0,p) ) |g(p)| dp < \infty
\label{Eq:gBound}
\end{equation}
and such that
\begin{equation}
\left| F(Q,J) \right| + \left| \frac{\partial F}{\partial Q}(Q,J) \right| \leq P(J),\qquad
(Q,J)\in S^1\times (0,\infty).
\label{Eq:FBound}
\end{equation}
Then, the key properties (a), (b) and (c) are satisfied.
\end{lemma}

\proof First, we note that the bound~(\ref{Eq:FBound}) implies that
\begin{equation}
|\hat{F}_k(J)| \leq \frac{1}{\sqrt{2\pi}}\int\limits_0^{2\pi} |F(Q,J)| dQ\leq \sqrt{2\pi} P(J),
\end{equation}
for all $k\in \Integer$. Together with the estimate~(\ref{Eq:PsiHatkBound}) this implies that
\begin{equation}
\int\limits_0^\infty |\hat{F}_k(J)| |\hat{\Psi}_k(J)| dJ
\leq \int\limits_0^\infty P(J(x_0,p)) \left[ |g(p)| + |g(-p)| \right] dp
 = \int\limits_{-\infty}^\infty P(J(x_0,p)) |g(p)| dp < \infty,
\end{equation}
where we have used the variable substitution $J = J(x_0,p)$ and the fact that $\omega(J) dJ = p dp$. Hence, it follows that $h_k := \hat{F}_k \hat{\Psi}_k^*\in L^1(0,\infty)$.

Next, by virtue of the Cauchy-Schwarz inequality one finds
\begin{equation}
\sum\limits_{k=-\infty}^\infty |\hat{F}_k(J)| 
 = \sum\limits_{k=-\infty}^\infty  \frac{1}{\sqrt{1 + k^2}} \sqrt{1 + k^2} |\hat{F}_k(J)|
 \leq C\left( \sum\limits_{k=-\infty}^\infty (1 + k^2) |\hat{F}_k(J)|^2 \right)^{1/2},
\end{equation}
with
\begin{equation}
C := \left( \sum\limits_{k=-\infty}^\infty  \frac{1}{1 + k^2} \right)^{1/2} < \infty.
\end{equation}
Using Parseval's identity and the hypothesis~(\ref{Eq:FBound}), this yields 
\begin{equation}
\sum\limits_{k=-\infty}^\infty |\hat{F}_k(J)| \leq C \left[ 
\int\limits_0^{2\pi}\left( | F(Q,J) |^2 + \left| \frac{\partial F}{\partial Q}(Q,J) \right|^2 \right) dQ \right]^{1/2}
 \leq \sqrt{2\pi} C P(J),
\end{equation}
and thus using the estimate~(\ref{Eq:PsiHatkBound}) again,
\begin{equation}
\sum\limits_{k=-\infty}^\infty \int\limits_0^\infty |\hat{F}_k(J)| |\hat{\Psi}_k(J)| dJ
 \leq C\int\limits_{-\infty}^\infty P(J(x_0,p)) |g(p)| dp < \infty,
\end{equation}
which shows that property (a) holds. Property (b) is a consequence of Lemma~\ref{Lem:RiemannLebesgue} and the fact that $h_k\in L^1(0,\infty)$.

Finally, in order to verify property (c) we compute the absolute convergent series
\begin{eqnarray*}
&& \sum\limits_{k=-\infty}^\infty \hat{F}_k(J) \hat{\Psi}_k(J)^* e^{-i\omega(J) t}\\
 &=& \frac{\omega(J)}{p} \frac{1}{\sqrt{2\pi}} \sum\limits_{k=-\infty}^\infty \hat{F}_k(J) 
 \left[ g(p) e^{-i k[Q(x_0,p) + \omega(J) t]} + g(-p) e^{i k[Q(x_0,p) - \omega(J) t]} 
 \right]_{p = \sqrt{2(H(J) - V(x_0))}}\Theta\left[ H(J) - V(x_0) \right] \\
  &=&\frac{\omega(J)}{p} \left[ g(p) F_t(Q(x_0,p),J) + g(-p) F_t(-Q(x_0,p),J) 
  \right]_{p = \sqrt{2(H(J) - V(x_0))}}\Theta\left[ H(J) - V(x_0) \right],
\end{eqnarray*}
where we have used the Fourier series representation of the function $F(\cdot,J)$ and the definition of the function $F_t$ in the last step. Integrating both sides over $J$, using the variable substitution $J = J(x_0,p)$ again and noticing that $Q(x_0,p) = 2\pi - Q(x_0,-p)$ we obtain
\begin{equation}
\int\limits_0^\infty \sum\limits_{k=-\infty}^\infty \hat{F}_k(J) \hat{\Psi}_k(J)^* e^{-i\omega(J) t} dJ
 = \int\limits_{-\infty}^\infty g(p) F_t(Q(x_0,p), J(x_0,p)) dp
 = \int\limits_{-\infty}^\infty f(t,x_0,p) g(p) dp,
\end{equation}
which shows property (c) and concludes the proof of the lemma.
\qed

The proof of Theorem~\ref{Thm:PSM2} can now be completed easily. Since $f_0\in C^1_0(\Gamma)$ it follows that $F\in C^1(S^1\times (0,\infty))$ is continuously differentiable and bounded, and further $F(Q,J)$ must vanish identically for large enough values of $J$. Furthermore, using Eqs.~(\ref{Eq:JDef},\ref{Eq:QDef}) (or comparing the two representations of the Vlasov equation, Eqs.~(\ref{Eq:Vlasov},\ref{Eq:VlasovAA1D}), with each other)
one finds
\begin{equation}
\omega\frac{\partial F}{\partial Q} 
 = p\frac{\partial f_0}{\partial x} - V'(x)\frac{\partial f_0}{\partial p},
\end{equation}
which shows that $\partial F/\partial Q$ is bounded (and, in fact, converges to zero) in the vicinity of the equilibrium point $(x,p) = (x_0,0)$. Then, the function
\begin{equation}
P(J) := \max\limits_{Q\in S^1} 
\left[ \left| F(Q,J) \right| + \left| \frac{\partial F}{\partial Q}(Q,J) \right| \right],\qquad J > 0,
\end{equation}
is continuous, bounded and vanishes identically for large $J$, such that the hypotheses of Lemma~\ref{Lem:Fck} are fulfilled. This concludes the proof of Theorem~\ref{Thm:PSM2}.

{\bf Remark}: It is clear that the assumption of $f_0$ having compact support can be relaxed. For example, it is sufficient to require that $f_0\in C^1(\Gamma)$ satisfies a bound of the form
\begin{equation}
\left| f_0(x,p) \right| 
 + \frac{T(H(x,p))}{2\pi}
  \left| p\frac{\partial f_0}{\partial x}(x,p) - V'(x)\frac{\partial f_0}{\partial p}(x,p) \right|
  \leq \alpha e^{-\beta H(x,p)},\qquad
(x,p)\in \Gamma,
\label{Eq:f0Bound}
\end{equation}
for some constants $\alpha,\beta > 0$, then the statement of the theorem still holds for all continuous functions $g$ satisfying
\begin{equation}
\int\limits_{-\infty}^\infty |g(p)| e^{-\beta H(x_0,p)} dp < \infty.
\end{equation}

\subsection{Proof of Theorem~\ref{Thm:PSM4}}

The proof of Theorem~\ref{Thm:PSM4} proceeds along the same lines as its one-dimensional counterpart in the last subsection, so we only provide a sketch and mention the relevant differences. The Fourier transform of the test function yields
\begin{equation}
\hat{\Psi}_k(J) = \frac{1}{T(E,L)}\frac{1}{\sqrt{2(E-V_L(r_0))}} \frac{1}{r_0}
\left[ g(p_+) e^{-i k\cdot Q(x_0,p_+)} + g(p_-) e^{-i k\cdot Q(x_0,p_-)} \right]
\Theta\left[ E - V_L(r_0) \right]_{(E,L) = I^{-1}(J)},
\label{Eq:PsiHatk2D}
\end{equation}
where $I^{-1}$ is the inverse of the map $I$ defined in Eq.~(\ref{Eq:IMap}) and $p_\pm := (\pm\sqrt{2(E - V_L(r_0)}, L)$ are the two solutions of $J(x_0,p) = I(E,L)$ for given $(E,L)\in \Omega$. To derive this result we have used the fact that $dQ_r dQ_\varphi = \omega_r dr d\varphi/|p_r| = \omega_r d^2 p/(r_0|p_r|)$. Eq.~(\ref{Eq:PsiHatk2D}) implies the bound
\begin{equation}
| \hat{\Psi}_k(J)| \leq \frac{1}{T(E,L)}\frac{1}{\sqrt{2(E-V_L(r_0))}} \frac{1}{r_0}
\left[ |g(p_+)| + |g(p_-)| \right]\Theta\left[ E - V_L(r_0) \right]_{(E,L) = I^{-1}(J)},
\end{equation}
for all $k\in \Integer^d$ and all $J\in\Omega_J$. Hence, if $\hat{F}_k(J)$ satisfies a bound of the type
\begin{equation}
\sum\limits_{k\in \Integer^2} |\hat{F}_k(J)| \leq C_1 P(J),
\label{Eq:FBoundWish}
\end{equation}
with $C_1$ a constant and $P:\Omega_J\to \Real$ a continuous function, then it follows that
\begin{equation}
\sum\limits_{k\in \Integer^2} \int\limits_{\Omega_J} |\hat{F}_k(J)| |\hat{\Psi}_k(J)| d^2 J
 \leq \frac{C_1}{2\pi}\int\limits_{H(x_0,p) < \Phi_\infty} P(J(x_0,p)) |g(p)| d^2 p,
\label{Eq:FSBound}
\end{equation}
where we have used the variable substitution $J = J(x_0,p)$ and the fact that $d^2 J = (|p_r|/\omega_r) dp_r dp_\varphi = (r_0|p_r|/\omega_r) d^2 p$ which can be inferred using the Jacobi matrix in Eq.~(\ref{Eq:JacobiMatrix}). Thus, the properties (a), (b) and (c) follow again if the right-hand side of Eq.~(\ref{Eq:FSBound}) is finite and if $F$ satisfies the bound~(\ref{Eq:FBoundWish}). This, in turn can be guaranteed if $F\in C^2(T^2\times \Omega_J)$ with a bound of the form
\begin{equation}
\left| F(Q,J) \right|^2 + \sum\limits_{A,B=r,\varphi}
 \left| \frac{\partial^2 F}{\partial Q_A\partial Q_B}(Q,J) \right|^2 \leq P(J)^2,\qquad
(Q,J)\in T^2\times \Omega_J,
\label{Eq:FBound2}
\end{equation}
which implies
\begin{eqnarray}
\sum\limits_{k\in\Integer^2} |\hat{F}_k(J)| 
 &=& \sum\limits_{k\in\Integer^2}  \frac{1}{\sqrt{1 + |k|^4}} \sqrt{1 + |k|^4} |\hat{F}_k(J)|
\nonumber\\
 &\leq& C\left( \sum\limits_{k\in\Integer^2} (1 + |k|^4) |\hat{F}_k(J)|^2 \right)^{1/2}
\nonumber\\
 &\leq& C\left[ \int\limits_{T^2}\left( \left| F(Q,J) \right|^2
  + \sum\limits_{A,B=r,\varphi}
 \left| \frac{\partial^2 F}{\partial Q_A\partial Q_B}(Q,J) \right|^2 \right) d^2 Q \right]^{1/2}
 \nonumber\\
 &\leq& 2\pi C P(J),
\end{eqnarray}
with
\begin{equation}
C := \left( \sum\limits_{k\in\Integer^2}  \frac{1}{1 + |k|^4} \right)^{1/2} < \infty,
\end{equation}
and the desired bound~(\ref{Eq:FBoundWish}) follows with $C_1 = 2\pi C$. As in the one-dimensional case, the bound~(\ref{Eq:FBound2}) can be guaranteed by demanding $f_0\in C^2_0(\Gamma_{bound})$ and noting that
\begin{eqnarray}
\omega_r\frac{\partial}{\partial Q_r} &=& p_r\frac{\partial}{\partial r} 
 + \left( \frac{p_\varphi}{r^2} - \omega_\varphi \right)\frac{\partial}{\partial\varphi}
 - \left(\Phi'(r) - \frac{p_\varphi^2}{r^3}\right)\frac{\partial}{\partial p_r}
\nonumber\\
 &=& p_x\frac{\partial}{\partial x} + p_y\frac{\partial}{\partial y}
 - \frac{\Phi'(r)}{r}\left( x\frac{\partial}{\partial p_x} + y\frac{\partial}{\partial p_y} \right)
 - \omega_\varphi\frac{\partial}{\partial Q^\varphi},\\
\frac{\partial}{\partial Q_\varphi} &=& \frac{\partial}{\partial \varphi}
 = x\frac{\partial}{\partial y} - y\frac{\partial}{\partial x} 
  + p_x\frac{\partial}{\partial p_y} - p_y\frac{\partial}{\partial p_x},
\end{eqnarray}
which has regular coefficients as long as one keeps away from the center $r = 0$.

\subsection{An alternative proof of Theorem~\ref{Thm:PSM1}}

Before concluding this appendix it is illustrative to consider the following alternative proof of Theorem~\ref{Thm:PSM1} which is based on a Fourier transformation on the whole phase space (and not just the angle variables) and is adapted from section~3 in Ref.~\cite{cMcV11}. For this, we go back to the action-angle representation~(\ref{Eq:NFPsi}) of $N_\varphi(t)$ and use the non-degeneracy condition~(\ref{Eq:NonDegeneracyCond}) to label each energy curve $C(E)$ by their frequency $\omega$ instead of the action variable $J$. Accordingly, instead of the function $F$ in Eq.~(\ref{Eq:ExplicitSolution}) we consider the function $G: S^1\times\Real\to \Real$, defined by
\begin{equation}
G(Q,\omega) := \left\{ \begin{array}{ll}
F(Q,J)\left| \frac{d\omega}{dJ}(J) \right|^{-1}, & \omega_{min} < \omega < \omega_{max},\\
0, & \hbox{otherwise},
\end{array} \right.
\end{equation}
where $(\omega_{min},\omega_{max})$ denotes the image\footnote{In general, one can show that
$\lim_{J\to 0}\omega(J) = \omega_0 := \sqrt{V''(x_0)}$, but the limit of $\omega(J)$ for $J\to \infty$ might be larger or smaller than $\omega_0$, depending on the form of the potential, see Appendix~\ref{App:PeriodFunction} for explicit examples. However, the precise form of the image $(\omega_{min},\omega_{max})$ is irrelevant for the proof in this section.} of the frequency function $\omega: (0,\infty) \to \Real$. Since the transformation $(x,p)\mapsto (Q,J)$ is symplectic and hence volume-preserving it follows that
\begin{equation}
\int\limits_{-\infty}^\infty\int\limits_0^{2\pi} G(Q,\omega) dQ d\omega = \int\limits_0^\infty\int\limits_0^{2\pi} F(Q,J) dQ dJ =
\int\limits_{\Gamma} f_0(x,p) dx dp,
\end{equation}
such that $G\in L^1(S^1\times \Real)$ if and only if $f_0\in L^1(\Gamma)$ which is satisfied according to the hypothesis of the theorem. Likewise, let $\Pi: S^1\times\Real\to \Real$ be the representation of the test function $\varphi$ in terms of the coordinates $(Q,\omega)$, such  that $\Pi(Q(x,p),\omega(J(x,p))) = \varphi(x,p)$ for all $(x,p)\in \Gamma_0$ and $\Pi(Q,\omega) = 0$ for all $\omega\notin (\omega_{min},\omega_{max})$. Then, Eq.~(\ref{Eq:NFPsi}) can be rewritten as
\begin{equation}
N_\varphi(t) = \int\limits_\Gamma f(t,x,p)\varphi(x,p) dx dp
 = \int\limits_{-\infty}^\infty\int\limits_0^{2\pi} G_t(Q,\omega) \Pi(Q,\omega) dQ d\omega,
\label{Eq:NgPsi}
\end{equation}
with
\begin{equation}
G_t(Q,\omega) := G(Q - \omega t,\omega).
\label{Eq:FrequencyRepresentationBis}
\end{equation}
The advantage of parametrizing the distribution and test function in terms of the frequency $\omega$ instead of $J$, is that in this representation, Eq.~(\ref{Eq:FrequencyRepresentationBis}) is formally equivalent to the distribution function for a collisionless kinetic gas on a periodic interval, with $Q$ and $\omega$ corresponding to the position and velocity, respectively, of the particle. 

Next, denote by $\tilde{G}$ the Fourier transform of $G$, defined by
\begin{equation}
\tilde{G}(k,\eta) := \frac{1}{2\pi} \int\limits_{-\infty}^\infty\int\limits_0^{2\pi} G(Q,\omega)
 e^{-i k Q - i\omega\eta} dQ d\omega,\qquad
 k\in \Integer,\quad\eta\in\Real.
\end{equation}
In Fourier space, the relation $G_t(Q,\omega) = G(Q - \omega t,\omega)$ reads $\tilde{G_t}(k,\eta) = \tilde{G}(k,\eta + kt)$, which shows that the rotations on the energy surfaces are converted into translations of the frequencies $\eta$ corresponding to the angular velocity variable. According to the Riemann-Lebesgue lemma $\tilde{G_t}(k,\eta)$ converges pointwise to $0$ for all fixed $(k,\eta)$ with $k\neq 0$. Therefore,
\begin{equation}
\lim\limits_{t\to \infty} \tilde{G_t}(k,\eta) = \delta_{k0}\tilde{G}(0,\eta) =: \tilde{G}_\infty(k,\eta).
\end{equation}
The right-hand side is precisely equal to the Fourier-transform of the angle average of $G$. Using Parseval's identity one obtains, assuming sufficient regularity to justify the passing of the limit under the integral, 
\begin{equation}
\lim\limits_{t\to\infty} N_\varphi(t) =
 \lim\limits_{t\to\infty}\int\limits_{-\infty}^\infty \sum\limits_{k=-\infty}^\infty
 \tilde{G_t}(k,\eta)^*\tilde{\Pi}(k,\eta) d\eta
 = \int\limits_{-\infty}^\infty\sum\limits_{k=-\infty}^\infty
 \tilde{G}_\infty(k,\eta)^* \tilde{\Pi}(k,\eta) d\eta 
  = \int\limits_{-\infty}^\infty \int\limits_0^{2\pi} \langle G \rangle (Q,\omega) \Pi(Q,\omega) dQ d\omega.
\end{equation}
This proof can be generalized to the two-dimensional case, assuming the frequency function $\omega(J)$ is locally invertible.

%%%%%%%%%%%%%%%%%%%%%%%%%%%%%%%%%%%%%%%%%%%%
\section{Monotonicity properties of the period function}
\label{App:PeriodFunction}
%%%%%%%%%%%%%%%%%%%%%%%%%%%%%%%%%%%%%%%%%%%%
 
There is an extensive literature on the monotonicity of the period function $T(E)$ defined in Eq.~(\ref{Eq:PeriodFunction}), see for instance Refs.~\cite{sCdW86,cC87}. In this appendix we briefly review some of these results and variants thereof which are used in the body of the paper. We assume that $V: I\to \Real$ is a $C^\infty$-differentiable function on the interval $I := (a,b)$, $-\infty\leq a < b \leq \infty$, that $V$ has a unique, non-degenerate global minimum at $x_0\in I$, such that $V'(x_0) = 0$, $V''(x_0) > 0$, and further assume that $V'(x) < 0$ for all $a < x < x_0$ and $V'(x) > 0$ for all $x_0 < x < b$. Additionally, we assume $\lim\limits_{x\to a} V(x) = \lim\limits_{x\to b} V(x)\in (0,\infty]$ and define
\begin{equation}
E_0:=V(x_0),\qquad E_{max} := \lim\limits_{x\to a} V(x).
\end{equation}
Following~\cite{cC87} we introduce the function $h: I\to \Real$ defined by
\begin{equation}
h(x) := \sign(x-x_0)\sqrt{2(V(x) - E_0)},\qquad x\in I.
\label{Eq:hDef}
\end{equation}
The assumed properties of $V$ imply that $h: I\to \Real$ is a smooth, strictly monotonously increasing function whose image is $h(I) = (-\sqrt{2(E_{max} - E_0)},+\sqrt{2(E_{max} - E_0)})$ and which satisfies $h(x_0) = 0$ and $h'(x_0) = \omega_0 := \sqrt{V''(x_0)} > 0$. Denote by $H = h^{-1}: h(I)\to I$ its inverse and by $H'$ the derivative of $H$. Then, the variable substitution $x = H(\sqrt{2(E-E_0)}\sin\alpha)$, $-\pi/2 < \alpha < \pi/2$, leads to the following expression for the period function:
\begin{equation}
T(E) = 2\int\limits_{-\pi/2}^{\pi/2} H'(\sqrt{2(E-E_0)}\sin\alpha) d\alpha.
\label{Eq:TE}
\end{equation}
From this and the smoothness of the function $H$, it follows immediately that $T: (E_0,E_{max})\to \Real$ is a $C^\infty$-differentiable function satisfying
\begin{equation}
\lim\limits_{E\to E_0} T(E) = T_0 := \frac{2\pi}{\omega_0}.
\end{equation}
Two useful expressions for the derivative of $T(E)$ can be obtained from Eq.~(\ref{Eq:TE}). The first expression, derived in Ref.~\cite{sCdW86} by a different method,  is obtained by differentiating both sides of Eq.~(\ref{Eq:TE}) with respect to $E$ and substituting back $x = H(\sqrt{2(E-E_0)}\sin\alpha)$, which yields
\begin{equation}
(E - E_0)\frac{dT}{dE}(E) 
 = \int\limits_{x_1(E)}^{x_2(E)} \frac{R(x)}{V'(x)^2} \frac{dx}{\sqrt{2(E-V(x))}}, 
\label{Eq:dTE1}
\end{equation}
with the function
\begin{equation}
R(x) := V'(x)^3\left[ \frac{V(x) - E_0}{V'(x)^2} \right]'
 = V'(x)^2 - 2[V(x)-E_0] V''(x).
\end{equation}
The second expression, derived in Ref.~\cite{cC87}, uses integration by parts and the identity $h(x) h'(x) = V'(x)$ to obtain
\begin{equation}
\frac{dT}{dE}(E) = 2\int\limits_{-\pi/2}^{\pi/2} H'''(\sqrt{2(E-E_0)}\sin\alpha) \cos^2\alpha d\alpha,
\qquad
H'''(y) = \left. \frac{h(x)}{V'(x)^5} N(x) \right|_{x = H(y)},
\label{Eq:dTE2}
\end{equation}
where
\begin{eqnarray}
N(x) &:=& R'(x) V'(x) - 3R(x) V''(x) \nonumber\\
 &=& V'(x)^4\left[ \frac{V(x) - E_0}{V'(x)^2} \right]'' \nonumber\\
 &=& 2[V(x) - E_0] [3V''(x)^2 - V'(x) V'''(x) ] - 3V'(x)^2 V''(x).
\label{Eq:NDef}
\end{eqnarray}
From these expressions it follows that the period function $T$ is monotonously increasing whenever $R(x) > 0$ or $N(x) > 0$ for all $x\in I\setminus \{ x_0 \}$. (It is sufficient to verify one of these conditions.) Similarly, $T$ is monotonously decreasing whenever $R(x) < 0$ or $N(x) < 0$ for all $x\in I\setminus \{ x_0 \}$.

We end this appendix by providing two explicit examples in which the monotonicity of the period function can be established rigorously. In the first one the function $R(x)$ has a definite sign but $N(x)$ has not; in the second example it is $N(x)$ that has a definite sign whereas $R(x)$ is indefinite.

The first example is the harmonic potential with a quartic perturbation
\begin{equation}
V(x) = \frac{1}{2}\omega_0^2 x^2 + \frac{1}{4} k x^4,\qquad x\in\Real,
\end{equation}
with $\omega_0 > 0$ and $k\geq 0$. In this case one has $E_0 = 0$ and
\begin{equation}
R(x) = -\frac{k}{2} x^4( 3\omega_0^2 + k x^2)\leq 0;
\end{equation}
hence it follows from Eq.~(\ref{Eq:dTE1}) that $T$ is constant in the harmonic case $k=0$ (as expected) and monotonously decreasing in the anharmonic case $k > 0$.\footnote{In contrast to this,
$$
N(x) = -\frac{3k}{2}x^4[ \omega_0^4 - 4\omega_0^2 k x^2 - k^2 x^4 ],
$$
which does not have a fixed sign, so that the monotonicity of $T(E)$ cannot be directly inferred from Eq.~(\ref{Eq:dTE2}).}

As a second example, consider the potential
\begin{equation}
V(x) := \log(x) + \frac{1}{2x^2},\qquad x > 0,
\label{Eq:IsothermalRescaled}
\end{equation}
which corresponds to the isothermal potential in the presence of the centrifugal term (with the rescaling $r = |L| x/\sqrt{3}$), see section~\ref{Sec:DarkMatterHalos}. $V$ has a global minimum at $x=1$ where $E_0 = \frac{1}{2}$; hence
\begin{equation}
2[V(x) - E_0] = \frac{1}{x^2} g(x^2-1),\qquad g(z) := (z + 1)\log(z+1) - z,\quad z > -1,
\end{equation}
and we obtain
\begin{equation}
x^{10} N(x) = \left[ g(z)(z^2 - 2z + 12) + 3z^2(z-2) \right]_{z=x^2-1},\qquad x > 0.
\label{Eq:NIsothermal}
\end{equation}

\begin{lemma}
\label{Lem:Isothermal}
The function $N(x)$ defined by Eq.~(\ref{Eq:NIsothermal}) is strictly positive for all $x > 0$, $x\neq 1$. Consequently, the associated period function $T(E)$ is monotonously increasing.\footnote{On the other hand,
$$
x^6 R(x) = \left[ z^2 + g(z)(z-2) \right]_{z=x^2-1}
$$
does not have a fixed sign, since it is negative for $z$ close to $-1$ but positive for large $z$.
}
\end{lemma}

\proof We divide the proof into two parts, corresponding to $|z| < 1$ and $z \geq 1$, respectively. For $|z| < 1$ we use the expansion
\begin{equation}
\log(1 + z) = \sum\limits_{k=1}^\infty (-1)^{k-1}\frac{z^k}{k}
\end{equation}
and find
\begin{equation}
g(z) = z^2\sum\limits_{k=0}^\infty \frac{(-1)^k}{(k+1)(k+2)} z^k,\qquad |z| < 1.
\end{equation}
From this, one obtains easily
\begin{equation}
g(z) \geq z^2\left( \frac{1}{2} - \frac{z}{6} \right),
\end{equation}
such that
\begin{equation}
x^{10} N(x) \geq \frac{z^2}{2}\left[ \left( 1 - \frac{z}{3} \right)(z^2 - 2z + 12) + 6(z-2) \right]
 = \frac{z^4}{6}(5-z) > 0,\qquad |z| < 1, z\neq 0.
\end{equation}
On the other hand, for $z\geq 1$ one has $g'(z) = \log(z+1) > 0$ such that $g(z) \geq g(1) = 2\log(2)-1 =:\delta > 1/3$. Furthermore, $z^2 - 2z + 12 \geq 11$ and $3z^2(z-2) \geq -32/9$ for all $z\geq 1$; hence
\begin{equation}
x^{10} N(x) \geq 11\delta - \frac{32}{9} > 11\left( \delta - \frac{1}{3} \right) > 0,
\end{equation}
which completes the proof of the lemma.
\qed

%%%%%%%%%%%%%%%%%%%%%%%%%%%%%%%%%%%%%%%%%%%%
\section{Hessian of the area function}
\label{App:AreaFunction}
%%%%%%%%%%%%%%%%%%%%%%%%%%%%%%%%%%%%%%%%%%%%

In this appendix, we analyze the properties of the area function $A: \Omega\to \Real$ defined in Eq.~(\ref{Eq:AreaFunction}). We assume the same properties on the potential $\Phi$ as the ones spelled out at the beginning of section~\ref{SubSec:2DModel}, such that the effective potential $V_L(r)$ has for each $L\neq 0$ a unique global non-degenerate minimum at $r = r_0$, corresponding to the minimal energy $E_0(L) = V_L(r_0)$.

We base our analysis on the same method discussed in the previous appendix, the main difference being the dependency of $L$ that enters into most of the quantities. Therefore, let $I_L := \{ r > 0 : V_L(r) < \Phi_\infty \}$ and introduce the function $h_L : I_L\to \Real$ defined by (cf. Eq.~(\ref{Eq:hDef}))
\begin{equation}
h_L(r) := \sign(r - r_0)\sqrt{2(V_L(r) - E_0(L))},\qquad r\in I_L.
\label{Eq:hLDef}
\end{equation}
For each $L\neq 0$, the function $h_L$ is smooth, monotonously increasing and satisfies $h_L(r_0) = 0$, $h_L'(r_0) = \omega_0(L) := \sqrt{V_L''(r_0)} > 0$. Next, we introduce the function $H: {\cal U}\to (0,\infty)$ defined by
\begin{eqnarray}
&& {\cal U} := \{ (L,y)\in \Real^2 : L\neq 0, y\in h_L(I_L) \},
\nonumber\\
&& H(L,y) := h_L^{-1}(y),\qquad (L,y)\in {\cal U},
\end{eqnarray}
which is smooth in both $L$ and $y$ as can easily be verified. The variable substitution $r = H(L,\sqrt{2(E-E_0(L))}\sin\alpha)$, $-\pi/2 < \alpha < \pi/2$, leads to the following expression for the area function
\begin{equation}
A(E,L) = 4(E - E_0(L))\int\limits_{-\pi/2}^{\pi/2} 
\frac{\partial H}{\partial y}(L,\sqrt{2(E-E_0(L))}\sin\alpha)\cos^2\alpha d\alpha,\qquad
(E,L)\in \Omega.
\label{Eq:AE1}
\end{equation}
Using integration by parts this can also be rewritten in the simpler form
\begin{equation}
A(E,L) = 2\sqrt{2(E - E_0(L))}\int\limits_{-\pi/2}^{\pi/2} 
H(L,\sqrt{2(E-E_0(L))}\sin\alpha)\sin\alpha d\alpha,\qquad
(E,L)\in \Omega,
\label{Eq:AE2}
\end{equation}
from which it also becomes clear that $A: \Omega\to \Real$ is a smooth function. The first partial derivatives of $A$ yield (setting $U := 1/H$)
\begin{eqnarray}
\frac{\partial A}{\partial E}(E,L) &=& 2\int\limits_{-\pi/2}^{\pi/2} 
\frac{\partial H}{\partial y}(L,\sqrt{2(E-E_0(L))}\sin\alpha) d\alpha,
\qquad (E,L)\in \Omega,\\
\frac{\partial A}{\partial L}(E,L) &=& 2L\int\limits_{-\pi/2}^{\pi/2} 
\frac{\partial U}{\partial y}(L,\sqrt{2(E-E_0(L))}\sin\alpha) d\alpha,
\qquad (E,L)\in \Omega,
\end{eqnarray}
and they provide alternative expressions for the period $T(E,L)$ of the radial motion and the azimuthal phase shift during one such period, see Eqs.~(\ref{Eq:TDef},\ref{Eq:DPsiDef}). In deriving this result we have used the identity $H(L,h_L(r)) = r$ for all $L\neq 0$ and all $r\in I_L$, which, upon differentiation with respect to $L$ yields
\begin{equation}
\frac{\partial H}{\partial L}(L,y) = -\frac{\partial H}{\partial y}(L,y)
 \frac{1}{y}\left( \frac{L}{H(L,y)^2} - \frac{\partial E_0(L)}{\partial L} \right).
\end{equation}
(Note that the right-hand side is regular at $y = 0$ because $H(L,0) = r_0$ and $\partial E_0(L)/\partial L = L/r_0^2$.) Further differentiation leads to the following expressions for the second derivatives of $A$:
\begin{eqnarray}
\frac{\partial^2 A}{\partial E^2}(E,L) &=& 2\int\limits_{-\pi/2}^{\pi/2} 
\frac{\partial^2 H}{\partial y^2}(L,\sqrt{2(E-E_0(L))}\sin\alpha)
\frac{\sin\alpha d\alpha}{\sqrt{2(E- E_0(L))}},
\qquad (E,L)\in \Omega,
\label{Eq:AEE}\\
\frac{\partial^2 A}{\partial E\partial L}(E,L) &=& 2L\int\limits_{-\pi/2}^{\pi/2} 
\frac{\partial^2 U}{\partial y^2}(L,\sqrt{2(E-E_0(L))}\sin\alpha)
\frac{\sin\alpha d\alpha}{\sqrt{2(E- E_0(L))}},
\qquad (E,L)\in \Omega,
\label{Eq:AEL}\\
\frac{\partial^2 A}{\partial L^2}(E,L) &=& 2\int\limits_{-\pi/2}^{\pi/2} 
\frac{\partial U}{\partial y}(L,\sqrt{2(E-E_0(L))}\sin\alpha) d\alpha
\nonumber\\
 &-& \frac{2L^2}{3}\int\limits_{-\pi/2}^{\pi/2} 
\frac{\partial^2(U^3)}{\partial y^2}(L,\sqrt{2(E-E_0(L))}\sin\alpha)
\frac{\sin\alpha d\alpha}{\sqrt{2(E- E_0(L))}},
\qquad (E,L)\in \Omega.
\label{Eq:ALL}
\end{eqnarray}
Substituting back $r = H(L,\sqrt{2(E-E_0(L))}\sin\alpha)$ into these expressions gives
\begin{eqnarray}
(E - E_0(L))\frac{\partial^2 A}{\partial E^2}(E,L) 
 &=& \int\limits_{r_1(E,L)}^{r_2(E,L)} \frac{R_L(r)}{V_L'(r)^2}\frac{dr}{\sqrt{2(E - V_L(r))}},
\qquad (E,L)\in \Omega,
\label{Eq:AEEBis}\\
(E - E_0(L))\frac{\partial^2 A}{\partial E\partial L}(E,L) 
 &=&L\int\limits_{r_1(E,L)}^{r_2(E,L)} \frac{4V_L'(r)[V_L(r) - E_0(L)] -r R_L(r)}
 { r^3 V_L'(r)^2}\frac{dr}{\sqrt{2(E - V_L(r))}},
\qquad (E,L)\in \Omega,
\label{Eq:AELBis}\\
(E - E_0(L))\frac{\partial^2 A}{\partial L^2}(E,L) 
 &=& -2(E - E_0(L))\int\limits_{r_1(E,L)}^{r_2(E,L)} \frac{1}{r^2}\frac{dr}{\sqrt{2(E - V_L(r))}}
 \nonumber\\
  &-& L^2\int\limits_{r_1(E,L)}^{r_2(E,L)} \frac{8V_L'(r)[V_L(r) - E_0(L)] - r R_L(r)}
 {r^5 V_L'(r)^2}\frac{dr}{\sqrt{2(E - V_L(r))}},
\qquad (E,L)\in \Omega,
\label{Eq:ALLBis}
\end{eqnarray}
where we have defined
\begin{equation}
R_L(r) := V_L'(r)^2 - 2[V_L(r) - E_0(L))] V_L''(r).
\end{equation}
The formulas~(\ref{Eq:AEE},\ref{Eq:AEL},\ref{Eq:ALL}) or (\ref{Eq:AEEBis},\ref{Eq:AELBis},\ref{Eq:ALLBis}) can be used to compute the determinant $\det(D^2 A(E,L))$ required for the verification of the non-degeneracy condition~(\ref{Eq:NonDegeneracyCond2D}).

There are a few interesting examples for which the function $H(L,y)$ and the area function $A(E,L)$ can be computed explicitly, and we mention three such examples at the end of this appendix. The first example is the harmonic potential $\Phi(r) = \omega_0^2 r^2/2$ in which case $H(L,y) = ( y + \sqrt{y^2 + 4\omega_0 L})/(2\omega_0)$ and $A(E,L) = \pi(E/\omega_0 - L)$. Since $A(E,L)$ is a linear function, there is no mixing.

The second example is the isochrone potential (see sections 2.2.2 and 3.5.2 in Ref.~\cite{BinneyTremaine-Book}, and references therein) for which
\begin{equation}
\Phi_b(r) = -\frac{1}{b + \sqrt{b^2 + r^2}},\qquad r > 0,
\end{equation}
with $b > 0$ a positive parameter. Note that in the limit $b\to 0$ this potential reduces to the Kepler potential $\Phi_0(r) = -1/r$. It is not difficult to verify that this potential satisfies all the conditions listed at the beginning of section~\ref{SubSec:2DModel}. Furthermore, the function $h_L$ defined in Eq.~(\ref{Eq:hLDef}) is given by
\begin{equation}
y = h_L(r) = \frac{\sqrt{b^2 + r^2}- a}{\sqrt{a} r},
\qquad r\in I_L = \left(\frac{r_0^2}{2a},\infty \right),
\end{equation}
where we have defined $a := \sqrt{b^2 + r_0^2}$. Inverting the function $h_L$, one finds
\begin{equation}
H(L,y) = \frac{r_0^2}{\sqrt{r_0^2 + a b^2 y^2} - a^{3/2} y }
 = \frac{ \sqrt{r_0^2 + a b^2 y^2} + a^{3/2} y}{1 - ay^2},\qquad |y| < a^{-1/2}.
\label{Eq:HLIsochrone}
\end{equation}
Here, the minimum of the potential $E_0(L)$ and its location $r_0$ are determined by the equations
\begin{equation}
E_0(L) = -\frac{2}{\left( |L| + \sqrt{4b + L^2} \right)^2},\qquad
r_0^2 = \frac{1}{16}\left( |L| + \sqrt{4b + L^2} \right)^4 - b^2.
\end{equation}
Introducing Eqs.~(\ref{Eq:HLIsochrone}) into Eq.~(\ref{Eq:AE2}) yields the explicit expression\footnote{To compute the integral one uses the symmetry of the interval and the integral identity
$$
\int\limits_{-\pi/2}^{\pi/2} \frac{d\alpha}{1 - k\sin\alpha} = \frac{\pi}{\sqrt{1-k^2}},\qquad k^2 < 1,
$$
which can be derived by elementary methods based on the variable substitution $\alpha = 2\arctan(u)$.}
\begin{equation}
A_b(E,L) = 2\pi \left[ \frac{1}{\sqrt{-2E}} - \frac{1}{\sqrt{-2E_0(L)}} \right],\qquad
E_0(L) < E < 0.
\label{Eq:AreaFunctionIsochrone}
\end{equation}
for the area function. Interestingly, the period for the radial motion is independent of the value of $L$ and $b$, and given by the exact Kepler formula $T(E) = 2\pi/(-2E)^{3/2}$. In the Kepler limit for which $b=0$ and $1/\sqrt{-2E_0(L)} = |L|$ it follows that $A(E,L)$ is linear in $|L|$, and hence the determinant condition~(\ref{Eq:NonDegeneracyCond2D}) is violated everywhere and there is no mixing. In contrast to this, when $b > 0$, the determinant condition~(\ref{Eq:NonDegeneracyCond2D}) is satisfied for all $(E,L)\in \Omega$ and according to Theorems~\ref{Thm:PSM3} and \ref{Thm:PSM4} mixing takes place.

The third example for which $A(E,L)$ can be computed explicitly occurs for the potential\footnote{Note that for this potential the function $r\mapsto r^3\Phi_*'(r) = r + a_2$ does not converge to zero for $r\to 0$; still it has positive derivative and its image is $(a_2,\infty)$; hence the effective potential $V_L$ has a unique minimum for each $L^2 > a_2$.}
\begin{equation}
\Phi_*(r) = -\frac{1}{r} - \frac{a_2}{2r^2},\qquad r > 0,
\label{Eq:KeplerPert}
\end{equation}
with $a_2$ a real constant different from zero. In this case, the effective potential $V_L(r)$ is the same as in the Kepler case with the replacement $L^2\mapsto L^2 - a_2$, and hence
\begin{equation}
A_*(E,L) = 2\pi \left[ \frac{1}{\sqrt{-2E}} - \sqrt{L^2 - a_2} \right],\qquad 
-\frac{1}{2(L^2 - a_2)} < E < 0,\quad L^2 > a_2.
\end{equation}
Since $A_*(E,L)$ is a nonlinear function of both $E$ and $L$, the determinant condition~(\ref{Eq:NonDegeneracyCond2D}) is satisfied everywhere and mixing takes place.

%%%%%%%%%%%%%%%%%%%%%%%%%%%%%%%%%%%%%%%%%%%%
\section{Conditions on the metric coefficients and structure of the effective potential $V_{L,m}$}
\label{App:EffPotential}
%%%%%%%%%%%%%%%%%%%%%%%%%%%%%%%%%%%%%%%%%%%%

In this appendix we specify our precise conditions on the metric coefficient $K(r) := S^2(r) N(r)$ (which determines the norm of the Killing vector field $\partial/\partial_t$) belonging to the static, spherically symmetric spacetimes discussed in section~\ref{Sec:BlackHoles} and show that they imply a qualitative behavior for the effective potential
\begin{equation}
V_{L,m}(r) = K(r)\left( m^2 + \frac{L^2}{r^2} \right),
\end{equation}
describing free-falling particles of mass $m$ in that spacetime, which is similar to the Schwarzschild case. Following~\cite{eCoS15a} we assume that $K: (0,\infty)\to \Real$ is a smooth function satisfying the conditions
\begin{enumerate}
\item[(i)] $K(r)\to 1$, $r^2K'(r)\to 2M > 0$ for $r\to \infty$,
\item[(ii)] $K(r_H) = 0$ and $K'(r_H) > 0$ for some $r_H > 0$.
\item[(iii)] $K(r) > 0$ and $K'(r) > 0$ for all $r > r_H$.
\end{enumerate}
In particular, these conditions are satisfied for any static, spherically symmetric and asymptotically flat black hole solution of total mass $M$ with non-degenerate event horizon at $r = r_H$. We further assume:
\begin{enumerate}
\item[(iv)] The function $P: [r_H,\infty)\to \Real$ defined by
\begin{equation}
P(r) := \frac{2K(r) - rK'(r)}{r^3K'(r)},\qquad r\geq r_H
\end{equation}
has a unique critical point at $r = r_{ms}$, say.
\end{enumerate}
Note that according to the the assumptions (i)--(iii) the function $P$ is well-defined, is negative at $r = r_H$ and positive for large $r$, behaving as $P(r) \simeq (Mr)^{-1}(1 - 3M/r)$ for large $r$ (which is exact in the Schwarzschild case). Therefore, condition (iv) implies that $r = r_{ms}$ is a global maximum of $P$ and that $P$ has a unique zero $r_{ph}$ in the interval $(r_H,r_{ms})$ (the meaning of $r_{ph}$ and $r_{ms}$ will become clear further below). With these assumptions and notation we can draw the following conclusions about the effective potential. First, we note that $V_{L,m}(r_H) = 0$ while for large $r$, $V_{L,m}\simeq m^2(1 - 2M/r)$. For $L = 0$ the effective potential is clearly increasing for all $r > r_H$. For $L\neq 0$, we can write its derivative as
\begin{equation}
V_{L,m}'(r) = L^2K'(r)\left[ \frac{m^2}{L^2} - P(r) \right].
\end{equation}
This implies that $V_{L,m}(r)$ is strictly increasing for all $r > r_H$ as long as
\begin{equation}
L < L_{ms},\qquad L_{ms} := \frac{m}{\sqrt{P(r_{ms})}},
\end{equation}
while $V_{L,m}'$ has two zeroes $r_{max}(L),r_{min}(L)$ (corresponding, respectively, to a local maximum and minimum of $V_{L,m}$) on the interval $[r_H,\infty)$ for all $L > L_{ms}$. These zeroes are further restricted by the inequalities $r_{ph} < r_{max}(L) < r_{ms} < r_{min}(L)$. As $L$ increases from $L_{ms}$ to $\infty$, $r_{max}(L)$ decreases monotonically from $r_{ms}$ to $r_{ph}$ while $r_{min}(L)$ increases monotonically from $r_{ms}$ to $\infty$ monotonically. The radii $r_{ph}$ and $r_{ms}$ correspond, respectively, to circular photon orbits and to the innermost stable circular orbit which is also an inflection point of $V_{L,m}$. (In the Schwarzschild case one has $r_{ph} = 3M$, $r_{ms} = 6M$, $L_{ms} = \sqrt{12} M m$.)

The range of parameters characterizing bound orbits which is relevant for section~\ref{Sec:BlackHoles} can be formulated as follows:
\begin{equation}
L > L_{ms},\qquad 
E_{min}(L) := \sqrt{V_{L,m}(r_{min}(L))} \leq E < \min\{ \sqrt{V_{L,m}(r_{max}(L))} , m \}
 =: E_{max}(L),
\label{Eq:EminEmax}
\end{equation}
where the lower limit $E = E_{min}(L)$ corresponds to stable circular orbits of radius $r = r_{min}(L)$.

\bibliographystyle{unsrt}
\bibliography{../References/refs_kinetic}

\begin{thebibliography}{10}

\bibitem{jLoP73}
J.L. Lebowitz and O.~Penrose.
\newblock Modern ergodic theory.
\newblock {\em Phys. Today}, 26(2):23--29, 1973.

\bibitem{CornfeldFominSinai-Book}
I.P. Cornfeld, S.V. Fomin, and Ya.G. Sinai.
\newblock {\em Ergodic Theory}.
\newblock Springer-Verlag, New York, 1982.

\bibitem{dL62a}
D.~Lynden-Bell.
\newblock Stellar dynamics. only isolating integrals should be used in {J}eans
  theorem.
\newblock {\em Monthly Notices Roy Astronom. Soc.}, 124:1--9, 1962.

\bibitem{dL62}
D.~Lynden-Bell.
\newblock The stability and vibrations of a gas of stars.
\newblock {\em Monthly Notices Roy Astronom. Soc.}, 124:279--296, 1962.

\bibitem{dL67}
D.~Lynden-Bell.
\newblock Statistical mechanics of violent relaxation in stellar systems.
\newblock {\em Monthly Notices Roy Astronom. Soc.}, 136:101--121, 1967.

\bibitem{sTmHdL86}
S.~Tremaine, M.~H\'enon, and D.~Lynden-Bell.
\newblock {H}-functions and mixing in violent relaxation.
\newblock {\em Monthly Notices Roy Astronom. Soc.}, 219:285--297, 1986.

\bibitem{sT99}
S.~Tremaine.
\newblock The geometry of phase mixing.
\newblock {\em Monthly Notices Roy Astronom. Soc.}, 307:877--883, 1999.

\bibitem{gCrSmFbGpKpA14}
G.~N. Candlish, R.~Smith, M.~Fellhauer, B.~K. Gibson, P.~Kroupa, and
  P.~Assmann.
\newblock Phase mixing due to the galactic potential: steps in the position and
  velocity distributions of popped star clusters.
\newblock {\em Monthly Notices Roy Astronom. Soc.}, 437:3702--3717, 2014.

\bibitem{dM99}
D.~Merritt.
\newblock Elliptical galaxy dynamics.
\newblock {\em Publications of the Astronomical Society of the Pacific},
  111(756):129--168, February 1999.

\bibitem{pDeJmAeMdN17}
P.~Dom\'inguez, E.~Jim\'enez, M.~Alcubierre, E.~Montoya, and D.~N\'u{\~n}ez.
\newblock Description of the evolution of inhomogeneities on a dark matter halo
  with the {V}lasov equation.
\newblock {\em Gen.Rel.Grav.}, 49:123, 2017.

\bibitem{BinneyTremaine-Book}
J.~Binney and S.~Tremaine.
\newblock {\em Galactic Dynamics (Second Edition)}.
\newblock Princeton University Press, Princeton, New Jersey, 2008.

\bibitem{aAmC14}
A.~Akbarian and M.W. Choptuik.
\newblock Critical collapse in the spherically-symmetric {E}instein-{V}lasov
  model.
\newblock {\em Phys. Rev.}, D90(10):104023, 2014.

\bibitem{pRoS18}
P.~Rioseco and O.~Sarbach.
\newblock Phase space mixing in the equatorial plane of a {K}err black hole.
\newblock {\em Phys. Rev. D}, 98(12):124024, 2018.

\bibitem{cMcV11}
C.~Mouhot and C.~Villani.
\newblock On {L}andau damping.
\newblock {\em Acta Math.}, 207:29--201, 2011.

\bibitem{bY16}
B.~Young.
\newblock Landau damping in relativistic plasmas.
\newblock {\em J. Math. Phys.}, 57:021502, 2016.

\bibitem{rMeT17}
R.~Mathew and E.~Tiesinga.
\newblock Phase-space mixing in dynamically unstable, integrable few-mode
  quantum systems.
\newblock {\em Phys. Rev. A}, 96:013604, 2017.

\bibitem{tDaKeKyS02}
T.V. Dudnikova, A.~I. Komech, E.~A. Kopylova, and Y.~M. Suhov.
\newblock On convergence to equilibrium distribution, {I}. {T}he
  {K}lein-{G}ordon equation with mixing.
\newblock {\em Comm. Math. Phys.}, 225:1--32, 2002.

\bibitem{tDaKnRyS02}
T.V. Dudnikova, A.~I. Komech, N.E. Ratanov, and Y.~M. Suhov.
\newblock On convergence to equilibrium distribution, {II}. {T}he wave equation
  in odd dimensions, with mixing.
\newblock {\em J. Stat. Phys.}, 108:1219--1253, 2002.

\bibitem{cM19}
C.~Mitchell.
\newblock Weak convergence to equilibrium of statistical ensembles in
  integrable hamiltonian systems.
\newblock {\em J. Math. Phys.}, 60:052702 (1)--(15), 2019.

\bibitem{ReedSimon80}
M.~Reed and B.~Simon.
\newblock {\em Methods of Modern Mathematical Physics, Vol. II}.
\newblock Academic Press, San Diego, 1980.

\bibitem{Arnold-Book}
V.I. Arnold.
\newblock {\em Mathematical Methods of Classical Mechanics}.
\newblock Springer-Verlag, New York, 1989.

\bibitem{sCdW86}
S.-N. Chow and D.~Wang.
\newblock On the monotonicity of the period function of some second order
  equations.
\newblock {\em \^{C}asopis pro p\^{e}stov\'an\'i matematiky}, 111:14--25, 1986.

\bibitem{cC87}
C.~Chicone.
\newblock The monotonicity of the period function for planar {H}amiltonian
  vector fields.
\newblock {\em J. Diff. Eq.}, 69:310--321, 1987.

\bibitem{pRoS16}
P.~Rioseco and O.~Sarbach.
\newblock Accretion of a relativistic, collisionless kinetic gas into a
  {S}chwarzschild black hole.
\newblock {\em Class. Quantum Grav.}, 34(9):095007, 2017.

\bibitem{Zehnder-Book}
E.~Zehnder.
\newblock {\em Lectures on Dynamical Systems: Hamiltonian Vector Fields and
  Symplectic Capacities}.
\newblock European Mathematical Society, Zurich, 2010.

\bibitem{jNcFsW96}
J.~Navarro, C.S. Frenk, and S.D.M. White.
\newblock The structure of cold dark matter halos.
\newblock {\em Astrophysical Journal}, 462:563, 1996.

\bibitem{jNcFsW97}
J.~Navarro, C.S. Frenk, and S.D.M. White.
\newblock A universal density profile from hierarchical clustering.
\newblock {\em Astrophys. J.}, 490:493--508, 1997.

\bibitem{aB95}
A.~Burkert.
\newblock The structure of dark matter halos in dwarf galaxies.
\newblock {\em Astrophysical Journal}, 447:L25--L28, 1995.

\bibitem{pRoS17}
P.~Rioseco and O.~Sarbach.
\newblock Spherical steady-state accretion of a relativistic collisionless gas
  into a {S}chwarzschild black hole.
\newblock {\em J. Phys. Conf. Ser.}, 831(1):012009, 2017.

\bibitem{tHeF08}
T.~Hinderer and E.E. Flanagan.
\newblock {Two timescale analysis of extreme mass ratio inspirals in Kerr. I.
  Orbital Motion}.
\newblock {\em Phys. Rev. D}, 78:064028, 2008.

\bibitem{MoBoschWhite-Book}
H.~Mo, F.~van~den Bosch, and S.~White.
\newblock {\em Galaxy Formation and Evolution}.
\newblock Cambridge University Press, Cambridge, U.K., 2010.

\bibitem{bC68}
B.~Carter.
\newblock Global structure of the {K}err family of gravitational fields.
\newblock {\em Phys. Rev.}, 174:1559--1571, 1968.

\bibitem{mWrP70}
M.~Walker and R.~Penrose.
\newblock On quadratic first integrals of the geodesic equations for type [22]
  spacetimes.
\newblock {\em Commun.Math.Phys.}, 18:265--274, 1970.

\bibitem{hA11}
H.~Andr{\'{e}}asson.
\newblock The {E}instein-{V}lasov system/kinetic theory.
\newblock {\em Living Reviews in Relativity}, 14(4), 2011.

\bibitem{dFjJjS15}
D.~Fajman, J.~Joudioux, and J.~Smulevici.
\newblock A vector field method for relativistic transport equations with
  applications.
\newblock {\em Analysis \& PDE}, 10:1539--1612, 2017.

\bibitem{dFjJjS17}
D.~Fajman, J.~Joudioux, and J.~Smulevici.
\newblock The stability of the {M}inkowski space for the {E}instein-{V}lasov
  system.
\newblock 2017.
\newblock arXiv:1707.06141.

\bibitem{mT16}
M.~Taylor.
\newblock The global nonlinear stability of {M}inkowski space for the massless
  {E}instein--{V}lasov system.
\newblock {\em Ann. PDE}, 3:9, 2017.

\bibitem{hLmT20}
H.~Lindblad and M.~Taylor.
\newblock Global stability of {M}inkowski space for the {E}instein--{V}lasov
  system in the harmonic gauge.
\newblock {\em Arch. Ration. Mech. Anal.}, 235:517--633, 2020.

\bibitem{lBdFjJjSmT20}
L.~Bigorgne, D.~Fajman, J.~Joudioux, J.~Smulevici, and M.~Thaller.
\newblock Asymptotic stability of {M}inkowski space-time with non-compactly
  supported massless {V}lasov matter.
\newblock 2020.
\newblock arXiv:2003.03346.

\bibitem{lApBjS18}
L.~Andersson, P.~Blue, and J.~Joudioux.
\newblock Hidden symmetries and decay for the {V}lasov equation on the {K}err
  spacetime.
\newblock {\em Comm. Partial Differential Equations}, 43:47--65, 2018.

\bibitem{lB20}
L.~Bigorgne.
\newblock Decay estimates for the massless {V}lasov equation on {S}chwarzschild
  spacetimes.
\newblock 2020.
\newblock arXiv:2006.03579.

\bibitem{eCoS15a}
E.~Chaverra and O.~Sarbach.
\newblock Radial accretion flows on static, spherically symmetric black holes.
\newblock {\em Class. Quantum Grav.}, 32:155006, 2015.

\end{thebibliography}
%%%%%%%%%%%%%%%%%%%%%%%%%%%%%%%%%%%%%%%%%%%%

\end{document}